\documentclass[letterpaper,11pt]{article}
\usepackage{graphics,graphicx}
\usepackage{longtable}
\usepackage{wrapfig}
\usepackage{amsmath,verbatim,amssymb,epsfig,color,lscape}
\usepackage{natbib}
\usepackage{multirow,multicol}
\usepackage[pagestyles]{titlesec}
\usepackage[normalem]{ulem}
\usepackage{bm}
\usepackage{url}
\usepackage{setspace}
\usepackage{multirow}

\newcommand{\bi}{\begin{itemize}}
\newcommand{\ei}{\end{itemize}}

\makeatletter

\usepackage[outerbars,color]{changebar}
\ifx\pdfoutput\undefined
\else\ifnum\pdfoutput>0
  \usepackage{pdfcolmk}
\fi\fi
\cbcolor{black}

\usepackage[margin=1.0in]{geometry}
\newfont{\rmm}{cmr10 at 11pt}
\rmm

\title{Stochastic Approximation Hamiltonian Monte Carlo}

\begin{document}
\title{
Stochastic Approximation Hamiltonian Monte Carlo
}
\author{
Jonghyun Yun, \ Minsuk Shin, \ Ick Hoon Jin\thanks{
Address for all correspondence.
Yun is an Independent Scholar, Mansfield, TX, 76063. USA
Shin is Assistant Professor, Department of Statistics, 
University of South Carolina, Columbia, SC. 29201. USA.
Jin is Assistant Professor, Department of Applied Statistics,
Yonsei University, Seoul. 03722. Republic of Korea. 
Liang is Professor, Department of Statistics, Purdue University, West Lafayette, IN 47907. USA.
}, \ and \ Faming Liang
}
\date{}
\maketitle
\bibliographystyle{Chicago}

\begin{abstract}
  Recently, the Hamilton Monte Carlo (HMC) has become widespread as one of the
  more reliable approaches to efficient sample generation processes. However,
  HMC is difficult to sample in a multimodal posterior distribution because the
  HMC chain cannot cross energy barrier between modes due to the energy
  conservation property. In this paper, we propose a Stochastic Approximate
  Hamilton Monte Carlo (SAHMC) algorithm for generating samples from multimodal
  density under the Hamiltonian Monte Carlo (HMC) framework. SAHMC can
  adaptively lower the energy barrier to move the Hamiltonian trajectory more
  frequently and more easily between modes.
% The convergence of the algorithm is established under mild conditions. 
% Gaussian mixture model and neural network model show that
  Our simulation studies show that the potential for SAHMC to explore a
  multimodal target distribution more efficiently than HMC based implementations.
\end{abstract}
\hspace*{1cm}{\bf Keywords}: 
Hamiltonian Monte Carlo; Stochastic approximation Monte Carlo; multimodality; \\
\hspace*{2.9cm}Scalable computation; Neural Networks

\section{Introduction}

Markov chain Monte Carlo (MCMC) is one of the conventional algorithms for
Bayesian inference to generate samples from the posterior distribution. However,
many proposed MCMC algorithms are very inefficient due to randomness in sample
generation. Recently, HMC \citep{Duane:87, Neal:10}, which imitates Hamiltonian
dynamics in the sample generation procedure, has gained popularity in the MCMC
communities as one of the efficient sampling methods \citep{Carpenter:17}.

To construct the MCMC using Hamiltonian dynamics, first define a Hamiltonian
function for the posterior probability distribution for which you want to
generate a sample. The Hamiltonian function consists of two variables: the
"position" variable corresponding to the realization from the posterior
probability distribution, and the auxiliary "momentum" variable, which helps the
chain to produce efficient samples using energy conservation characteristics.
Typically, an independent Gaussian distribution \citep{Neal:10} or Riemannian
manifold (RMHMC: Fisher information) \citep{Girolami:11} are assumed for the
momentum variable. By updating the momentum variable, a new state can be
proposed by calculating the trajectories moving along the same set of levels
according to the energy conservation properties. The state proposed by
Hamiltonian mechanics can be far from the current state, but nonetheless it is
likely to accept the sample.

HMC can be classified as one of the auxiliary variable MCMC algorithms and is
called {\it hybrid Monte Carlo} because it combines MCMC and the deterministic
simulation algorithm, the leapfrog method for implementing Hamiltonian dynamics.
The tuning parameters of the leapfrog method, leapfrog step-size, $\epsilon$,
and the number of leapfrog, $L$ are one of the practical issues in implementing
the HMC. The no-U-turn sampler (NUTS) has been proposed by \citet{Hoffman:14} to
eliminate the need to set the HMCs the number of leapfrog, by stopping
trajectory automatically when it starts to double back and retrace its steps. A
detailed description of HMC based on geometric foundations is given in
\citet{Betancourt:14} and the theoretical properties of HMC have been studied in
\citet{livingstone2016geometric} and \citet{Durmus:17}.

One of the well-known problems in HMC is the generation of samples from
multimodal posterior distributions. HMC and its variants \citep{Girolami:11, Hoffman:14, Neal:10, Shahbaba:13, Sohl-Dickstein:14} 
cannot easily move from one mode to another within a small number of iterations because the energy barrier between modes 
is high when the parameters of interest have different modes separated by low probability areas.

Several variants of the HMC algorithm have been proposed to solve the
multimodality problem of HMC. Wormhole Hamiltonian Monte Carlo (WHMC, \citet{Lan:2014})
modifies the RMHMC \citep{Girolami:11} to create a short path (a so-called
wormhole) that connects modes to facilitate movement between modes. This method
is useful if detailed and accurate knowledge about modes are provided, but such
requirement is not practical in high-dimensional multimodal distributions.
To resolve this issue, WHMC employs a mode discovering mechanism. According to
\cite{fok_optimization_2017}, however, this mechanism still requires some knowledge
about the target distribution; otherwise it may unstabilize WHMC.

A tempering approach is combined with HMC in \citet{Nishimura:16} and
the authors suggest a new type of mass matrix to lower the energy barrier
between modes so that the trajectory of Hamiltonian mechanics can move more
often from one mode to the other. However, this algorithm must specify an
appropriate temperature schedule so that the trajectory can efficiently navigate
the parameter space. In addition, the standard integrator is not applicable to
this sampler because the velocity of this sampler is unbounded in the area of
the low probability region. Last, this algorithm uses a non-volume preserving
integrator, therefore it requires to calculate the determinant of the Jacobian for the acceptance probability.

In this article, we propose a new algorithm, Stochastic Approximation
Hamiltonian Monte Carlo (SAHMC), for generating samples from a multimodal
density under the framework of the HMC. SAHMC is an HMC version of the
Stochastic Approximation Monte Carlo (SAMC) algorithm \citep{Liang:07ML,
  Liang:07}, which draws samples from each subregions with a predetermined
frequency. SAHMC use weights in SAMC, which are updated proportionate to the
differences between actual number of visits of subregions with the prespecified
frequency using stochastic approximation \citep{Robbins:51}, to adaptively lower
the energy barrier between modes and allow chains to easily pass through low
probability areas between modes. The convergence of the algorithm is established
under mild conditions. Compared to \citet{Lan:2014}, SAHMC does not need to know
the location of the mode before implementation. Compared to
\citet{Nishimura:16}, the specification of neither temperature schedule nor
variable-step integrators is required for SAHMC implementations. Numerical
results show that the new algorithm works well, especially if the target density
is a high-dimensional multimodal distribution.

The rest of the article is organized as follows. 
In Section 2, we describe the SAMC algorithm. 
In Section 3, we incorporate HMC under the framework of SAMC and study its theoretical property.
In Section 4, we test our SAHMC algorithm to the Gaussian mixture models along with extensive comparison with HMC.
In Section 5, we apply our SAHMC to neural network model and compare results with HMC.
In Section 6, we conclude the article with brief discussions.

\section{SAMC Algorithm}

Suppose that we are interested in sampling from the distribution
\begin{equation}\label{eq:fx}
 f({\bf x}) = c \psi({\bf x}), \quad {\bf x} \in {\cal X}, 
\end{equation}
where $c$ is an unknown normalizing constant and ${\cal X}$ is the sample space. 
For mathematical convenience, we assume that ${\cal X}$ is either finite or compact. 

We let $E_1$, $\cdots$, $E_m$ denote $m$ partition of ${\cal X}$ 
according to the potential energy function, $U({\bf x}) = -\log \psi({\bf x})$, 
i.e., $E_1 = \big\{{\bf x}: U({\bf x}) \leq u_1, {\bf x} \in {\cal X}\big\}$, 
$E_2 = \big\{{\bf x}: u_1 < U({\bf x}) \leq u_2, {\bf x} \in {\cal X}\big\}$, $\cdots$,
$E_{m-1} = \big\{{\bf x}: u_{m-2} < U({\bf x}) \leq u_{m-1}, {\bf x} \in {\cal X}\big\}$, 
and $E_m = \big\{{\bf x}: U({\bf x}) > u_{m-1}, {\bf x} \in {\cal X}\big\}$,
where $u_1 < u_2 < \cdots < u_{m-1}$ are pre-specified numbers. 
Let $\boldsymbol\pi = (\pi_1, \cdots, \pi_m)$ be an $m$-vector 
with $0 < \pi_i < 1$ and $\sum_{i=1}^m \pi_i = 1$, 
and denote the desired sampling frequency for each of the subregions. In
general, $\pi$'s are chosen to be uniform when there are no prior knowledge
availalbe about
$\psi(\bf x)$. 
Then, the estimate of equation (\ref{eq:fx}) can be written as
\begin{equation}\label{eq:fx_hat}
f_{\omega}(x) \propto \sum_{i=1}^m  \frac{\pi_i \psi({\bf x})}{\omega_i} I\big({\bf x} \in E_i\big).
\end{equation}
where the partition of normalizing constant, $\omega_i = \int_{E_i} \psi({\bf x}) dx$.
SAMC allows the existence of empty subregions in simulations and
provides an automatic way to learn the normalizing constants $\omega_1$, $\cdots$, $\omega_m$. 

Let $\{a_t\}$ denote the gain factor sequence which is positive, non-increasing sequence satisfying
\begin{equation}\label{eq:gain1}
\mbox{(a)} \quad \sum_{t=1}^{\infty} a_t = \infty, \qquad \mbox{and} 
\qquad \mbox{(b)} \quad \sum_{t=1}^{\infty}a_t^{\zeta} < \infty.
\end{equation}
for some $\zeta \in (1, 2)$. For example, \citep{Liang:07} suggests
\begin{equation}\label{eq:gain2}
a_t = \frac{t_0}{\max(t_0, t)}, \qquad t = 1, 2, \cdots
\end{equation}
for some specified value of $t_0 > 1$. 

Let $\theta_i^{(t)}$ denote a working estimate of $\log(\omega_i/\pi_i)$
obtained at iteration $t$, and let $\boldsymbol\theta^{(t)} = (\theta_1^{(t)},\ldots,\theta_m^{(t)})$.
With the foregoing notation, one iteration of SAMC can be described as follows:
\vskip .2cm

\underline{\bf SAMC Algorithm}
\begin{enumerate}
\item {\bf (Sample Generation)} Simulate a sample ${\bf x}^{(t+1)}$ by a single
  MH update, of which the invariant distribution is a working estimate of
  $f_{\omega}(x)$ at iteration $t$.

\item {\bf ($\theta$-updating step)}
  Set
\[ \boldsymbol\theta^{*} = \boldsymbol\theta^{(t)} + a_{t+1}\Big({\bf e}_t - \boldsymbol\pi\Big),\]
where ${\bf e}_t = \big(e_{t,1}, \cdots, e_{t,m}\big)$ and 
$e_{t,i} = 1$ if ${\bf x}_{t} \in E_i$ and 0 otherwise.
If $\boldsymbol\theta^{*} \in \Theta$, set $\boldsymbol\theta^{(t+1)} = \boldsymbol\theta^{*}$;
otherwise, set $\boldsymbol\theta^{(t+1)} = \boldsymbol\theta^{*} + {\bf c}^{*}$, 
where ${\bf c}^* = \big(c^*, \cdots, c^*\big)$ can be an arbitrary vector which satisfies the condition
$\boldsymbol\theta^{*} + {\bf c}^{*} \in \Theta$.
\end{enumerate}

For effective implementation of SAMC, several issues must be considered \citep{Liang:07}:
\begin{itemize}
\item {\it Sample Space Partition} \quad The sample space are partitioned
according to our goal and the complexity of the given problem. 
For example, when we generate samples from the distribution, 
the sample space can be partitioned according to the energy function. 
The maximum energy difference in each subregion should be bounded,  
for example, \citet{Liang:07} suggests to use 2. 
Within the same subregion, the behavior of the SAHMC move reduces to the local
HMC.

\item {\it Choice of the desired sampling distribution} \quad
If we aim to estimate $\boldsymbol\omega$, 
then we may set the desired distribution to be uniform, 
as is done in all examples in this article. 
However, we may set the desired distribution biased to low-energy regions. 
To ensure convergence, the partition of all sample spaces must be visited in proportion to the desired sampling distribution.

\item {\it Choice of $t_0$ and the number of iterations} \quad
To estimate $\boldsymbol\omega$, $\alpha_t$ should be very close to 0 at the end of simulations
and the speed of $\alpha_t$ going to 0 can be controlled by $t_0$. 
In practice, we choose $t_0$ according to the complexity of the problem; 
The more complex the problem is, the larger the value of $t_0$ that should be chosen. 
A large $t_0$ will make SAHMC reach all subregions quickly, even in the presence of multiple local energy minima.
\end{itemize}

\section{SAHMC Algorithm}

To substitute the sample generation step in SAMC to HMC, we at first define the potential energy function 
as
\begin{equation}\label{eq:2}
  U({\bf x}) = -\log \psi({\bf x})
\end{equation}
and the kinetic energy function as
\begin{equation}\label{eq:kin} 
K({\bf y}) = \frac{d}{2}\log(2\pi) + \frac{1}{2}\log |{\bf M}| + \frac{1}{2} {\bf y}{\bf M}^{-1}{\bf y}, 
\end{equation}
where the auxiliary variable ${\bf y}$ is interpreted as a momentum variable,  
$d$ is the dimension of ${\bf x}$, and covariance matrix ${\bf M}$ denotes a mass matrix. 
We can prespecify the mass matrix $M$ using a diagonal matrix
or define it using the Riemannian manifold \citep{Girolami:11}...
Then, the Hamiltonian and its corresponding probability function are 
\begin{equation}\label{eq:Ham1}
H({\bf x}, {\bf y}) = U({\bf x}) + K({\bf y}) \qquad \mbox{and} \qquad 
g({\bf x}, {\bf y}) \propto \exp \Big\{-U\big({\bf x}\big) - K\big({\bf y}\big)\Big\}
\end{equation}
The partial derivatives of $H({\bf x}, {\bf y})$ determines 
how ${\bf x}$ and ${\bf y}$ change over time, according to Hamiltonian equation,
\begin{equation}\label{eq:Ham2}\begin{split}
\dot{{\bf x}} &= ~~~\bigtriangledown_{{\bf y}} H({\bf x}, {\bf y}) = ~~~{\bf M}^{-1}{\bf y}\\
\dot{{\bf y}} &= -\bigtriangledown_{{\bf x}} H({\bf x}, {\bf y}) = -\bigtriangledown_{{\bf x}} U({\bf x})
\end{split}\end{equation}
Note that ${\bf M}^{-1}{\bf y}$ can be interpreted as velocity.

Under the aforementioned energy partition, the estimate of equation (\ref{eq:fx}) can be written as
\begin{equation}\label{eq:fx_hat}
f_{\omega}(x) \propto \sum_{i=1}^m \frac{\pi_i\psi({\bf x})}{\omega_i} I\big({\bf x} \in E_i\big)
= \sum_{i=1}^m \int_{\cal Y} \frac{\pi_i g({\bf x}, {\bf y})}{\omega_i} I\big({\bf x} \in E_i\big) dy.
\end{equation}
where the partition of normalizing constant, $\omega_i$, is
\begin{equation}\label{eq:nc}
\omega_i = \int_{E_i} \psi({\bf x}) dx = \int_{E_i}\int_{\cal Y} g({\bf x}, {\bf y}) dy dx. 
\end{equation}
SAHMC allows the existence of empty subregions in simulations and
provides an automatic way to learn the normalizing constants $\omega_1$, $\cdots$, $\omega_m$. 

Let $\theta_t^{i}$ denote a working estimate of $\log(\omega_i^t/\pi_i)$ obtained at iteration $t$ 
where $\omega_i^t$ is the $\omega_i$ value at iteration $t$, 
and 
\[ f_{\theta_t^i}(x) \propto \frac{\psi({\bf x})}{\exp(\theta_t^i)} I\big({\bf x} \in E_i\big)
  = \int_{\cal Y} \frac{g({\bf x}, {\bf y})}{\exp(\theta_t^i)} I\big({\bf x} \in E_i\big) dy. \]

With the foregoing notation, one iteration of SAHMC can be described as follows: 
\vskip .2cm

\underline{\bf SAHMC Algorithm}
\begin{enumerate}
\item {\bf (Momentum Updating)} Draw an independent normal random variable ${\bf y} \sim N_d\big(0, {\bf M}\big)$, 
and set ${\bf y}_t = {\bf y}$ and $K_t = K\big({\bf y}_t\big)$. 
Also, set ${\bf x}_t = {\bf x}$ and $U_t = U\big({\bf x}_t\big)$.

\item {\bf (Proposal Step)}
\begin{enumerate}
\item Set ${\bf x}_t^0 = {\bf x}_t$. Make a half step for the momentum at the beginning with 
\[ {\bf y}_t^0 \leftarrow {\bf y}_t - \frac{\epsilon}{2} 
\times \frac{\partial U({\bf x}_t)}{\partial {\bf x}} \bigg]_{{\bf x}_t^0}. \]
\item Alternate full steps for position and momentum. For $i = 1, \cdots, L-1$, do the following
 \begin{enumerate}
 \item Make a full step for the position: 
 ${\bf x}_t^i \leftarrow {\bf x}_t^{i-1} + \epsilon \times {\bf y}_t^{i-1}$.
 \item Make a full step for the momentum, except at the end of trajectory:
 \[ {\bf y}_t^i \leftarrow {\bf y}_t^{i-1} - \epsilon 
   \times \frac{\partial U({\bf x}_t^i)}{\partial {\bf x}} \bigg]_{x_t^i}. \]
\end{enumerate}

\item Make a half step for momentum at the end:
 \[ {\bf y}_t^L \leftarrow {\bf y}_t^{L-1} - \frac{\epsilon}{2} 
 \times \frac{\partial U({\bf x}_t^L)}{\partial {\bf x}} \bigg]_{{\bf x}_t^L}. \]
 \item Set negative momentum at the end of trajectory to make the proposal symmetric: 
 ${\bf y}_* = -{\bf y}_t^L$. Also, set ${\bf x}_* = {\bf x}_t^L$.
\end{enumerate}

\item {\bf (Decision Step)} Set $U_* = U({\bf x}_*)$ and $K_* = K({\bf y}_*)$, calculate
\begin{equation} 
r = \exp\Big\{\theta_t^{J(U_t)} - \theta_t^{J(U_*)}\Big\}\exp\Big\{U_t + K_t - U_* - K_*\Big\},
\end{equation}
where $J(U_t)$ denotes the index of the subregion that ${\bf x}_t$ belongs to.
Accept the proposal with probability $\min(1, r)$. If accepted, set ${\bf x}_{t+1} = {\bf x}_*$; 
otherwise ${\bf x}_{t+1} = {\bf x}_t$.

\item {\bf ($\theta$-updating step)} Set
\[ \boldsymbol\theta_{*} = \boldsymbol\theta_t + a_{t+1}\Big({\bf e}_t - \boldsymbol\pi\Big),\]
where ${\bf e}_t = \big(e_{t,1}, \cdots, e_{t,m}\big)$ and 
$e_{t,i} = 1$ if ${\bf x}_{t} \in E_i$ and 0 otherwise.
If $\boldsymbol\theta_{*} \in \Theta$, set $\boldsymbol\theta_{(t+1)} = \boldsymbol\theta_{*}$;
otherwise, set $\boldsymbol\theta_{(t+1)} = \boldsymbol\theta_{*} + {\bf c}^{*}$, 
where ${\bf c}^* = \big(c^*, \cdots, c^*\big)$ can be an arbitrary vector which satisfies the condition
$\boldsymbol\theta_{*} + {\bf c}^{*} \in \Theta$.
\end{enumerate}
%Since HMC satisfies the local positive condition; that is, for every $x \in {\cal X}$, 
%there exist $\epsilon_1 > 0$ and $\epsilon_2 > 0$ such that
%\begin{equation}
%||{\bf x} - {\bf y}|| \leq \epsilon_1 
%\quad \Rightarrow \quad
%q\big({\bf x}, {\bf y}\big) \geq \epsilon_2
%\end{equation}
%where $||z||$ denotes the norm of the vector $z$, 

In general, $\Theta$ is chosen to be a large compact set (e.g.
$[-10^{100},10^{100}]^m$), which is practically equivalent to
$\mathbb{R}^{m}$. Thanks to the location invariance of the target
distribution, a choice of a constant vector $\bf{c}^{*}$ does not affect the
theoretical convergence of SAHMC algorithm.

Like SAMC \citep{liang2009improving, Liang:07}, SAHMC also falls into the category 
of stochastic approximation algorithms \citep{Andrieu:05, Benveniste:90}, and the theoretical convergence results are provided in the Appendix.
The theory states that under mind conditions, we have 
\begin{equation}\label{eq:samc_thm}
\theta_t^{i} \to \left\{\begin{array}{lc}
C + \log\bigg(\int_{E_i} \int_{\cal Y}g\big({\bf x}, {\bf y}\big) dydx\bigg) - \log\Big(\pi_i + \nu\Big), & \mbox{if}~ E_i \neq \phi\\
-\infty & \mbox{if}~ E_i = \phi
\end{array}\right.
\end{equation}
where $\nu = \sum_{j \in \{i: E_i = \phi\}} \pi_j / (m - m_0)$ 
and $m_0$ is the number of empty subregions, and $C$ represents an arbitrary constant.
Since $f_{\boldsymbol\theta_t}({\bf x}) =
\sum\nolimits_{i=1}^{m}f_{\theta_t^i}(\bf x)$ is invariant with respect to a location transformation of $\boldsymbol\theta_t$,
$C$ cannot be determined by the SAHMC samples.
To determine the value of $C$, extra information is needed; 
for example, $\sum_{i=1}^m e^{\theta_t^{i}}$ is equal to a known number.

The main advantage of the SAHMC algorithm is that it can adaptively lower the energy barrier between modes 
and move the Hamiltonian trajectory more frequently and easily across the low
probability regions between modes. In the meanwhile, the HMC trajectory cannot move to another
mode beyond the energy barrier \citep{Nishimura:16}.

The energy barrier with
respect to $U({\bf x})$ from a position ${\bf x}_1$ to ${\bf x}_2$, is the
minimum amount of kinetic energy $K({\bf x})$ to reach ${\bf x}_2$ from ${\bf
  x}_1$ in a single iteration \citep{Nishimura:16}:
\begin{equation}\label{eq:barrier}
  B_{H}\Big({\bf x}_1, {\bf x}_2; U\Big) = \inf_{\boldsymbol\gamma \in C^0} \bigg\{ \max_{0 \leq t \leq 1} 
  U\big(\boldsymbol\gamma(t)\big) - U\big({\bf x}_1\big) \mid
  \boldsymbol\gamma\big(0\big) = {\bf x}_1 ~ \mbox{and} ~ \boldsymbol\gamma\big(1\big) = {\bf x}_2 \bigg\}
\end{equation} 
where $C^0$ denotes a class of continuous function. Note that due to the energy conservation property, 
\begin{equation}\label{eq:egp}
  U\big({\bf x}_t\big) - U\big({\bf x}_1\big) = K\big({\bf x}_1\big) - K\big({\bf x}_t\big) \leq K\big({\bf x}_1\big).
\end{equation}
If the kinetic energy of $K\big({\bf x}_1\big)$ is less than the energy barrier $B\big({\bf x}_1, {\bf x}_2; U\big)$, 
then HMC will not be able to reach ${\bf x}_2$.
% Suppose there are two modes, ${\bf x}_1$ and ${\bf x}_2$, in our target density and
% these two modes are separated by the low probability region. 
% In addition, assume the current location of the chain is near ${\bf x}_1$.
% For HMC, from equation (\ref{eq:egp}), the kinetic energy of $K({\bf x}_1)$ 
% should be larger than the energy barrier $B_H\big({\bf x}_1, {\bf x}_2; U\big)$
% to make the chain reach the other peak ${\bf x}_2$.
However, for SAHMC, the equation (\ref{eq:barrier}) can be rewritten as
\begin{equation}\label{eq:rebarrier}
B_{SA}\Big({\bf x}_1, {\bf x}_2; U\Big) = \inf_{\boldsymbol\gamma \in C^0} \bigg\{ \max_{0 \leq t \leq 1} 
U\big(\boldsymbol\gamma(t)\big) - U\big({\bf x}_1\big)
+ \Big(\theta^{J(U_t)} - \theta^{J(U_1)}\Big) \bigg|
\boldsymbol\gamma\big(0\big) = {\bf x}_1 ~ \mbox{and} ~ \boldsymbol\gamma\big(1\big) = {\bf x}_2 \bigg\}
\end{equation} 
%\begin{equation}\label{eq:reegp}
%\begin{split}
%U\big({\bf x}_t\big) - U\big({\bf x}_1\big) - \Big(\theta_t^{J(U_t)} - \theta_t^{J(U_1)}\Big) 
%&= K\big({\bf x}_1\big) - K\big({\bf x}_t\big) - \Big(\theta_t^{J(U_t)} - \theta_t^{J(U_1)}\Big)\\
%&\leq K\big({\bf x}_1\big) - \Big(\theta_t^{J(U_t)} - \theta_t^{J(U_1)}\Big) 
%\end{split}\end{equation}
where $J(U_t)$ and $J(U_1)$ denote the index of the subregions 
that ${\bf x}_{\boldsymbol\gamma(t)}$ and ${\bf x}_1$ belong to, respectively. 
From our assumption that the chain  currently stays near ${\bf x}_1$ for several iterations. 
That means the sample of $J(U_1)$ is oversampled than $\pi_{J(U_1)}$, 
while the sample of $J(U_t)$ is undersampled than $\pi_{J(U_t)}$, 
resulting in $\theta^{J(U_1)}$ being larger than $\theta^{J(U_t)}$. 
Then, under the SAHMC framework, the energy barrier can be lowered by $\theta^{J(U_t)} - \theta^{J(U_1)}$, 
so kinetic energy in $K\big({\bf x}_1\big)$ can move the trajectory more easily than in other modes ${\bf x}_2$. 
The amount of energy barriers lowered by SAHMC is determined adaptively according to the frequency of visits to the subregion. 

The other benefit of the SAHMC algorithm is its flexibility for other variants of the HMC. 
Because SAHMC can be implemented by adding one more step to the HMC, 
all existing HMC variants can be easily implemented under the SAHMC framework. 
For example, by replacing mass matrix ${\bf M}$ with Fisher information matrix, 
we can easily implement RMHMC \citep{Girolami:11} under the framework of SAHMC. 

% The no-U-turn sampler \citep{Hoffman:14} can also be easily implemented to SAHMC.

\section{Illustrative Examples}
\subsection{Gaussian Mixture Distributions \label{sec:gauss-mixt-distr}}

\begin{table}[htbp]
\caption{\label{tab:gmm}
Gaussian Mixture Examples: Comparison of SAHMC, HMC and NUTS}
\centering
\begin{tabular}{clrcrr} \hline\hline
\multicolumn{6}{l}{{\bf Set 1: a = -6, b = 4}}                                    \\
Parameter & Method & Time(s) & ESS (min, med, max) & s / min ESS & Relative Speed \\ \hline
$x_1$ 
          & HMC    & 30.6    & ( 787,  882,  941)  & 0.03888     & 1.00           \\
          & SAHMC  & 30.6    & (2041, 2984, 3549)  & 0.01499     & 2.59           \\ 
          & NUTS   & 30.0    & (32, 50, 225)       & 0.93750     & 0.04           \\ \hline
$x_2$
          & HMC    & 30.6    & ( 802,  879,  950)  & 0.03815     & 1.00           \\
          & SAHMC  & 30.6    & (2120, 3034, 3493)  & 0.01443     & 2.64           \\ 
          & NUTS   & 30.0    & (31, 50, 236)       & 0.96774     & 0.04           \\ \hline \hline
\multicolumn{6}{l}{{\bf Set 2: a = -8, b = 6}}                                    \\
Parameter & Method & Time(s) & ESS (min, med, max) & s / min ESS & Relative Speed \\ \hline
$x_1$
          & HMC    & 30.6    & ( 18,  25,   78)    & 1.70000     & 1.00           \\
          & SAHMC  & 30.6    & (533, 723, 1033)    & 0.05741     & 29.61          \\ 
          & NUTS   & 19.0    & (3, 3060, 9700)     & 6.33333     & 0.27           \\ \hline 
$x_2$
          & HMC    & 30.6    & ( 17,  26,  78)     & 1.80000     & 1.00           \\
          & SAHMC  & 30.6    & (581, 683, 825)     & 0.05267     & 34.18          \\ 
          & NUTS   & 19.0    & (3, 3479, 8841)     & 6.33333     & 0.28           \\ \hline \hline
\end{tabular}
\end{table}

As our illustrative example, we compare SAHMC with HMC and NUTS using the following Gaussian mixture distribution:
\begin{equation}\label{eq:gmm}
p({\bf x}) = \frac{1}{3} 
N_2\left[\left(\begin{array}{c} a       \\ a \end{array}\right), 
       \left(\begin{array}{cc} 1 & 0.9  \\ 0.9  & 1 \end{array}\right)\right] + 
\frac{1}{3} 
N_2\left[\left(\begin{array}{c} b       \\ b \end{array}\right), 
       \left(\begin{array}{cc} 1 & -0.9 \\ -0.9 & 1 \end{array}\right)\right] +
\frac{1}{3} 
N_2\Big[{\bf 0}, I_2\Big] 
\end{equation}
which is identical to that given by \citet{Gilks:98} 
except that the mean vectors are separated by a larger distance in each dimension. 
With this example, we show how SAHMC outperform 
to the original HMC method in generating samples from multimodal density. 
%using effective sample size and relative speeds.
%illustrate how SAHMC is able to generate sample from a multimodal density and 
%to show how much SAHMC can outperform to the original HMC method 
%in the aspect of effective sample size and relative speeds. 

NUTS improves upon HMC by automatically choosing optimal values for HMC's
tunable method parameters. It has been shown that NUTS samples complex distributions
effectively \citep{henderson_ti-stan_2019}.  NUTS can be easily implemented using
Stan software \cite{carpenter_stan_2017}
% (\url{https://mc-stan.org})
, which is available in many popular computing environments including R, Python, Julia, MATLAB, etc.

We used two sets of $(a, b)$; $(-6, 4)$ and $(-8, 6)$. 
For both HMC and SAHMC, we set the leapfrog step-size, $\epsilon = 0.3$, and
leapfrog steps, $L = 20$. We used the identity mass matrices of HMC and
SAHMC, so for fair comparison we used the
identity mass matrix for NUTS.
For NUTS implementation, the stepsize, $\epsilon$ is tuned using the
dual-averaging algorithm of \cite{Hoffman:14} so that the average
acceptance probability corresponds to a pre-specified value $\delta = 0.8$,
which is suggested as an optimal value by \citet{Hoffman:14}.
Other than the mass matrix, NUTS were implemented
using default parameter values 
in Stan.

To run SAHMC, we set the sample space ${\cal X} = [-10^{100}, 10^{100}]^2$ to be compact and
it was partitioned with equal energy bandwidth $\Delta u = 2$ into the following subregions: 
$E_1 = \{{\bf x}: -\log p({\bf x}) < 0\}$, $E_2 = \{{\bf x}: 0 \leq -\log p({\bf x}) < 2\}$, 
$\cdots$, and $E_{12} = \{{\bf x}: -\log p({\bf x}) > 20\}$.
Additionally, we set $t_0 = 5000$ and the desired sampling distribution, $\boldsymbol{\pi}$ to be uniform for SAHMC. 
All methods were independently run ten times and 
each run consists of 1,000,000 iterations, where the first 200,000 iterations 
were discarded as a burn-in process. 

Table \ref{tab:gmm} summarizes the performance of HMC, SAHMC and NUTS in aspects
to the effective sample sizes and relative speeds. HMC and SAHMC show no
differences in computational time; for both HMC and SAHMC, each run takes 30.6s
on a 2.6 GHz Intel i7 processor. NUTS automatically selects an appropriate
number of leapfrog steps in each iteration, so its computation time slightly
varies over each run. An average computation time of NUTS over ten runs are
reported in Table~\ref{tab:gmm}. Under Set 2, NUTS gives a high ESS when it
fails to visit all three modes. It is known that NUTS is likely to produce
approximately independent samples, in the sense of low autocorrelation. In this
example, however, NUTS's ESS is large only when it fails to fully explore target
distributions.

The relative speed, the computation time 
for generating one effective sample, of our SAHMC algorithm 
is about 2.6 times faster than HMC and 66 times faster than NUTS for Set 1 (a = -6, b = 4). 
For Set 2 (a = -8, b = 6), which has larger distance between modes 
and faces more difficulties in generating MCMC samples, 
the performance of SAHMC algorithm is approximately 30 times better than those
of HMC and 122 times better than those of NUTS.

\begin{figure}[htbp]
\caption{\label{fig:gmm}
Position and momentum coordinates for the first 1,000 iterations 
and scatter plots for the Gaussian mixture example with a = -8 and b = 6.
Row 1:   HMC (a-c). 
Row 2: SAHMC (d-f).
Row 3: NUTS (g-h).}
\centering
\begin{tabular}{ccc}
(a) HMC Position & (b) HMC Momentum & (c) Scatter Plot of HMC \\
\includegraphics[width=0.3\textwidth]{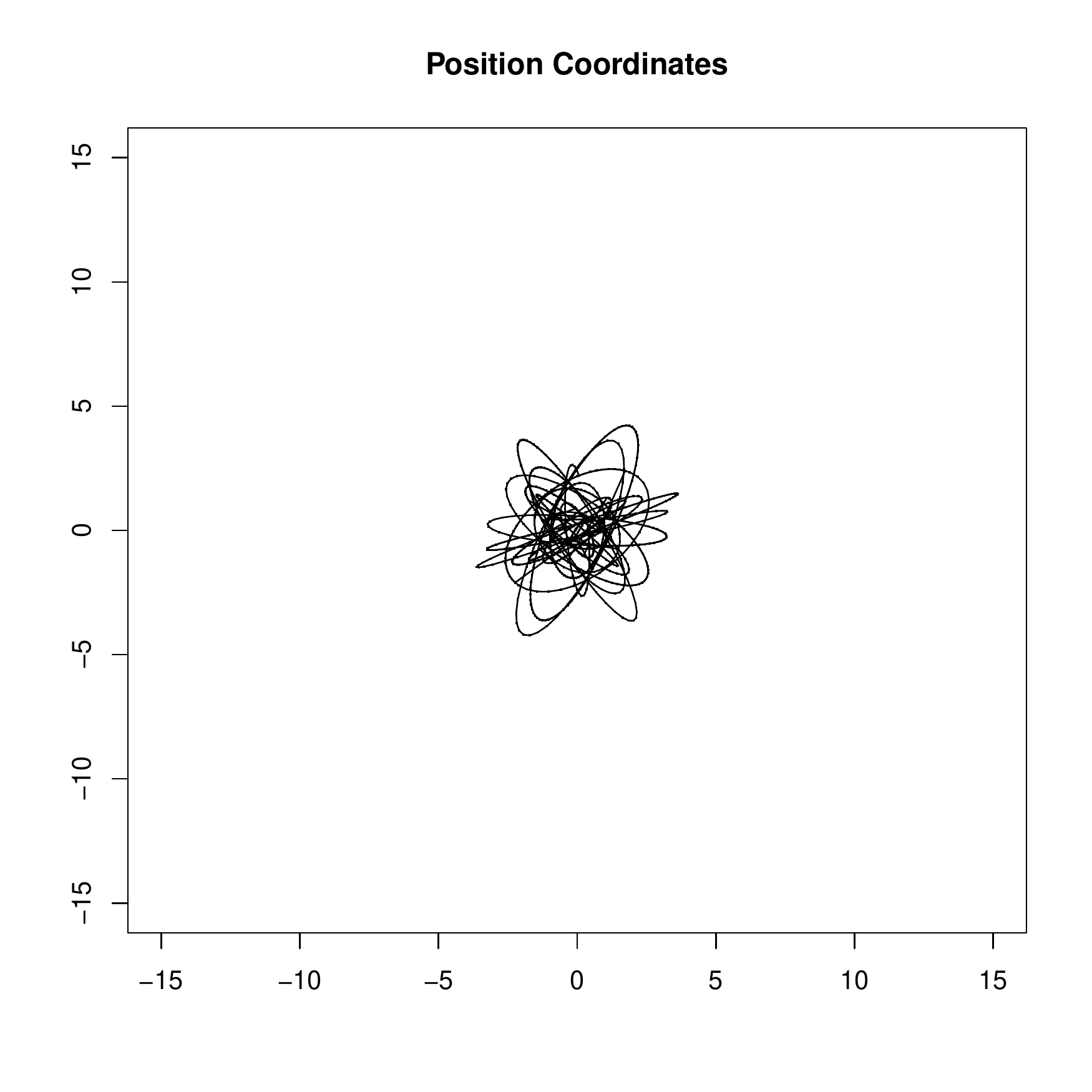} &
\includegraphics[width=0.3\textwidth]{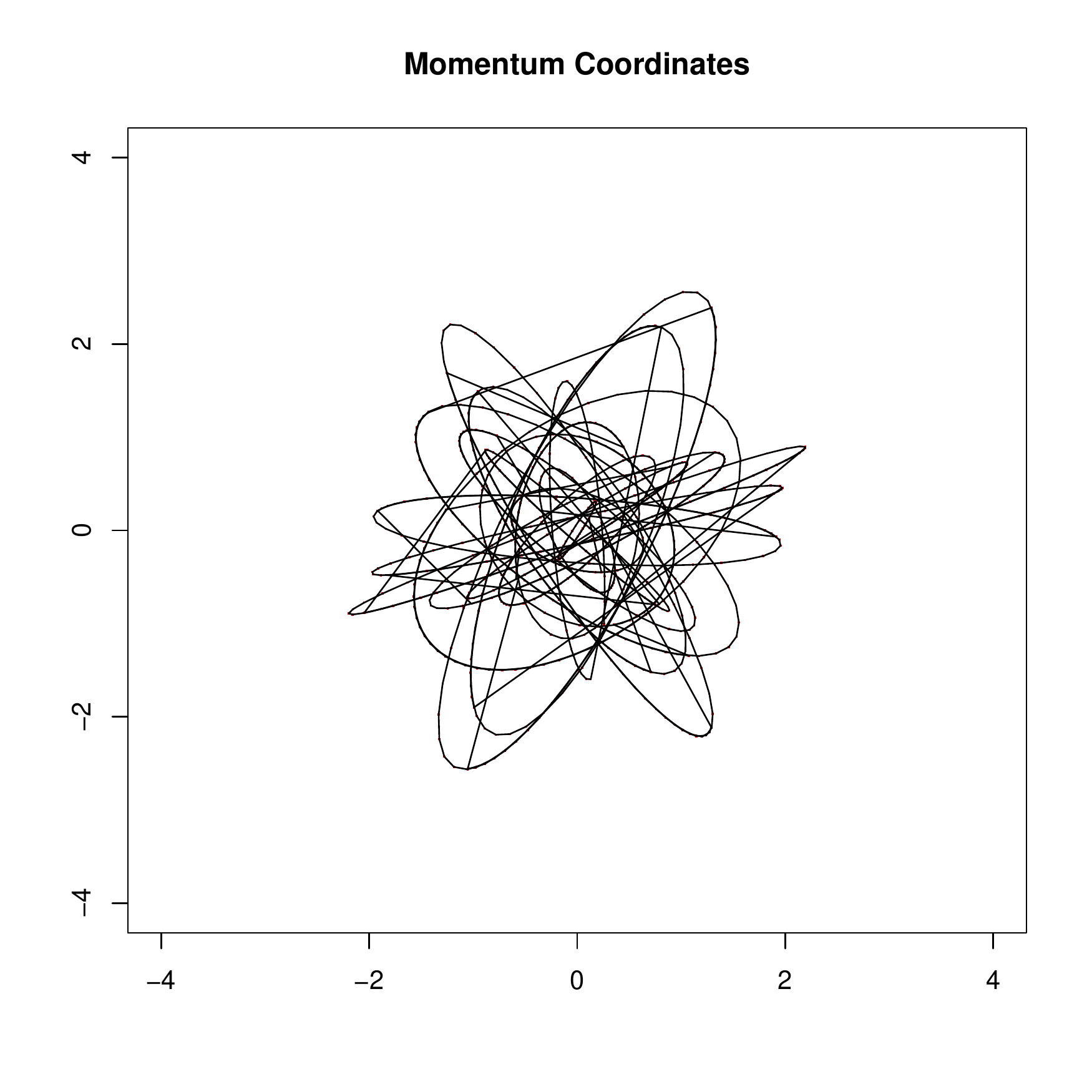} &
\includegraphics[width=0.3\textwidth]{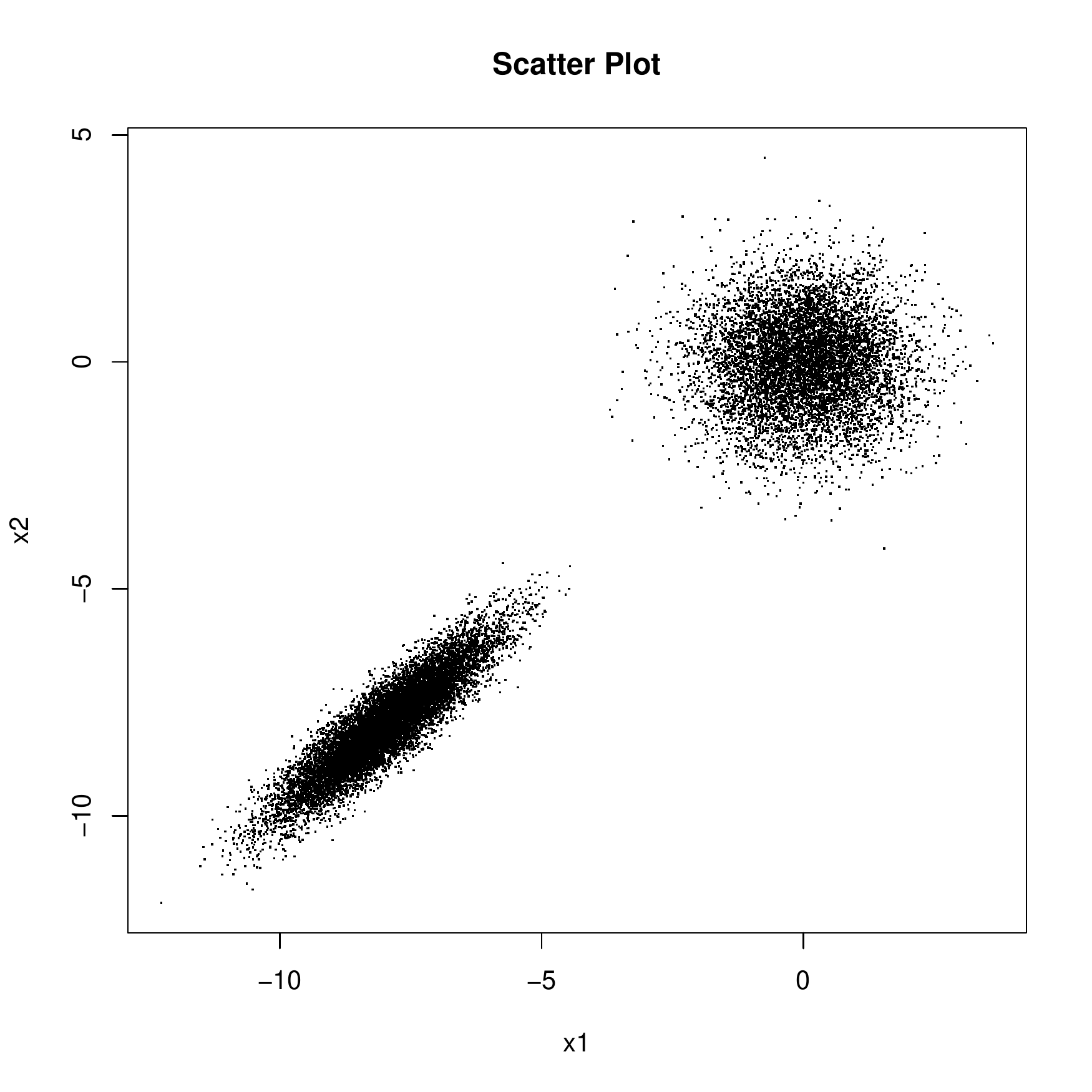} \\
(d) SAHMC Position & (e) SAHMC Momentum & (f) Scatter Plot of SAHMC \\
\includegraphics[width=0.3\textwidth]{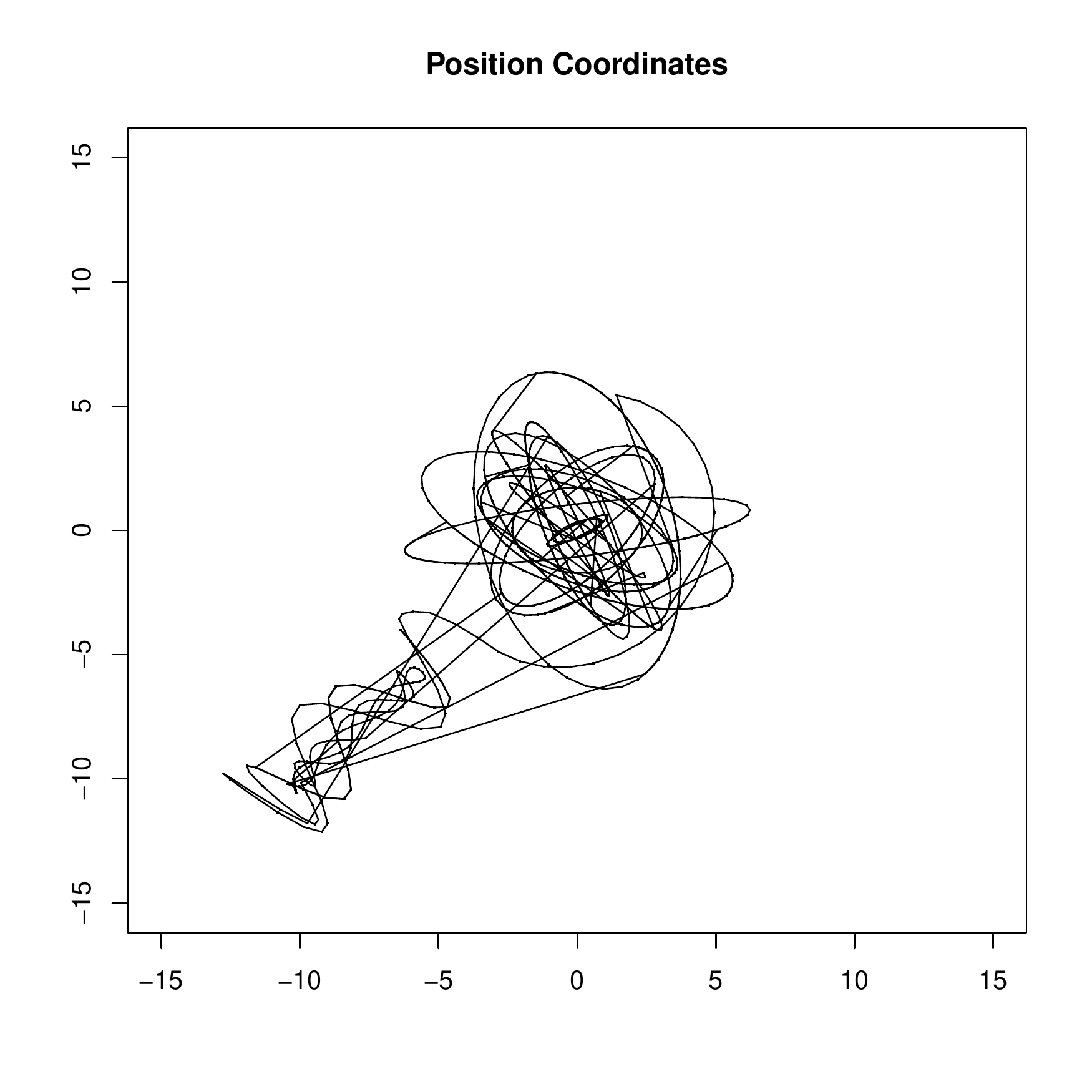} &
\includegraphics[width=0.3\textwidth]{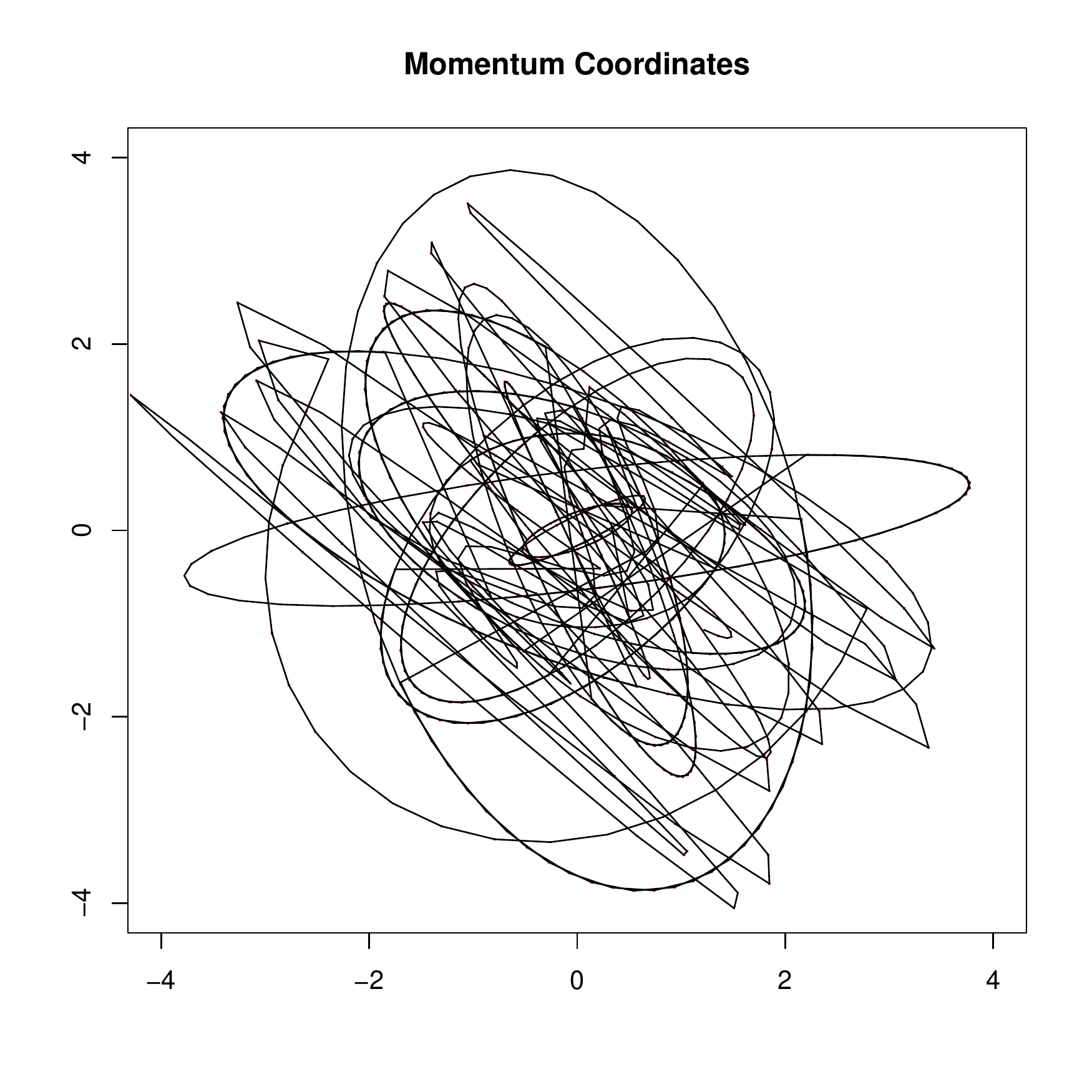} &
\includegraphics[width=0.3\textwidth]{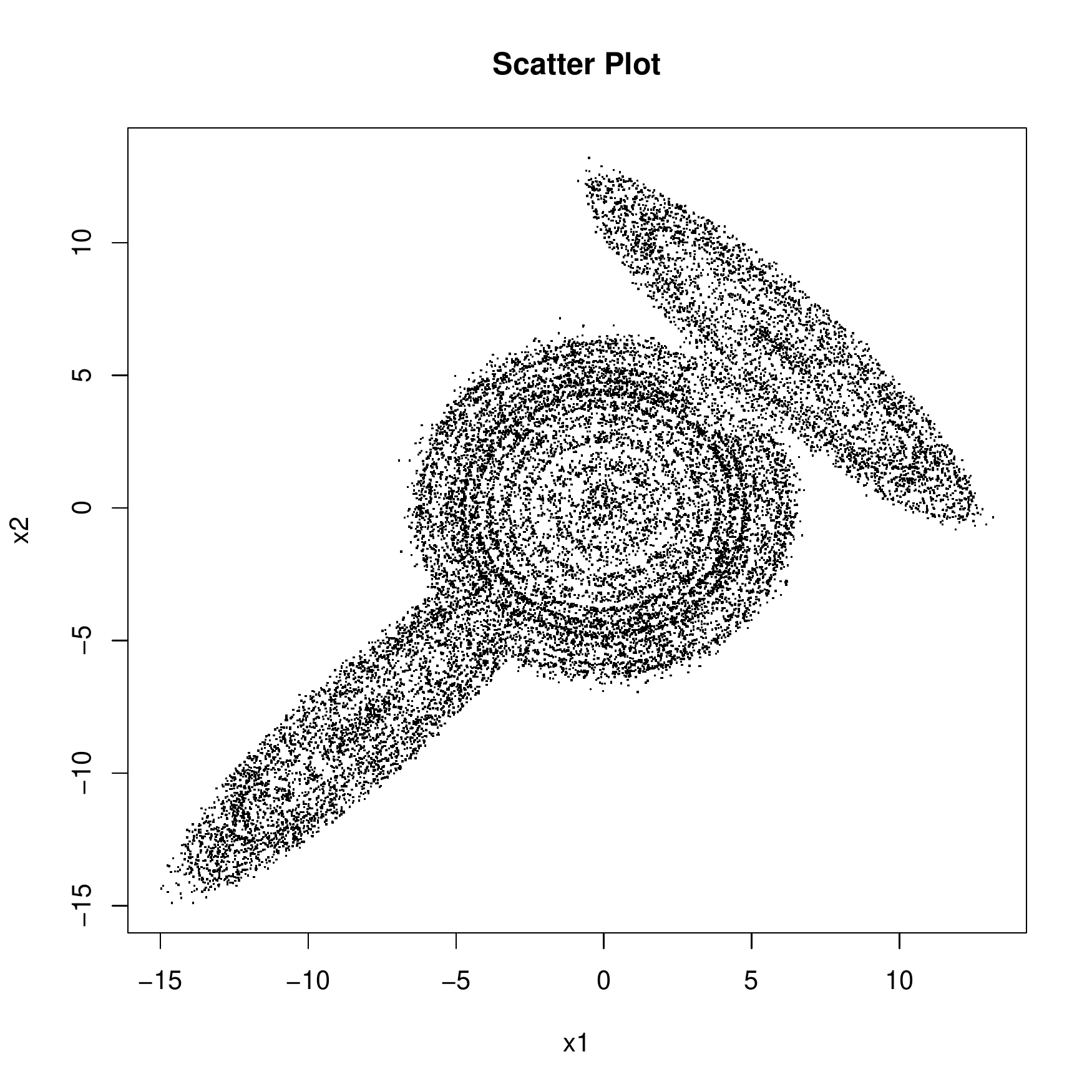} \\
  (g) Scatter Plot of NUTS (1st run) & (h) Scatter Plot of NUTS (2nd run) & (i) Scatter Plot of NUTS (3rd run) \\
 \includegraphics[width=0.3\textwidth]{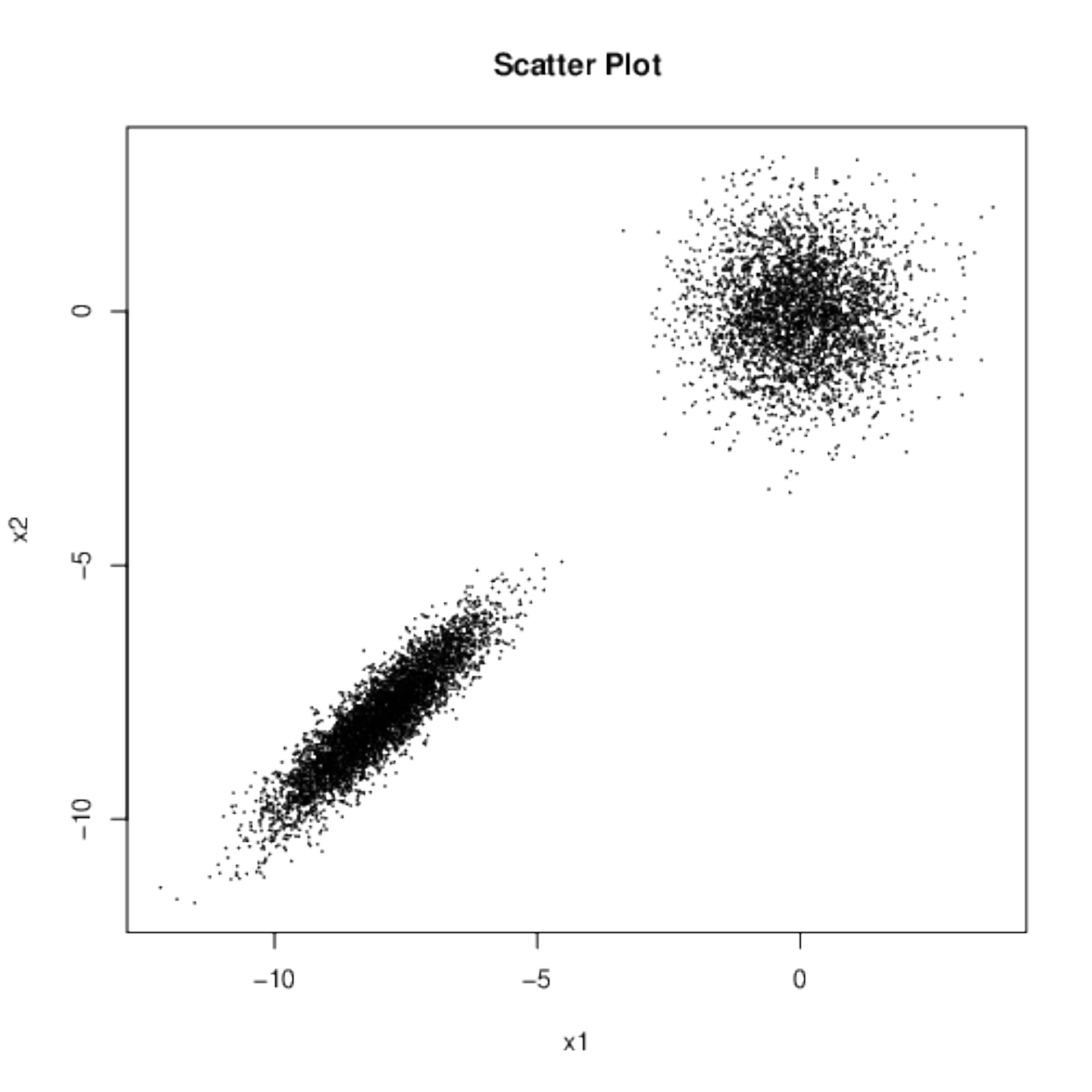} &
 \includegraphics[width=0.3\textwidth]{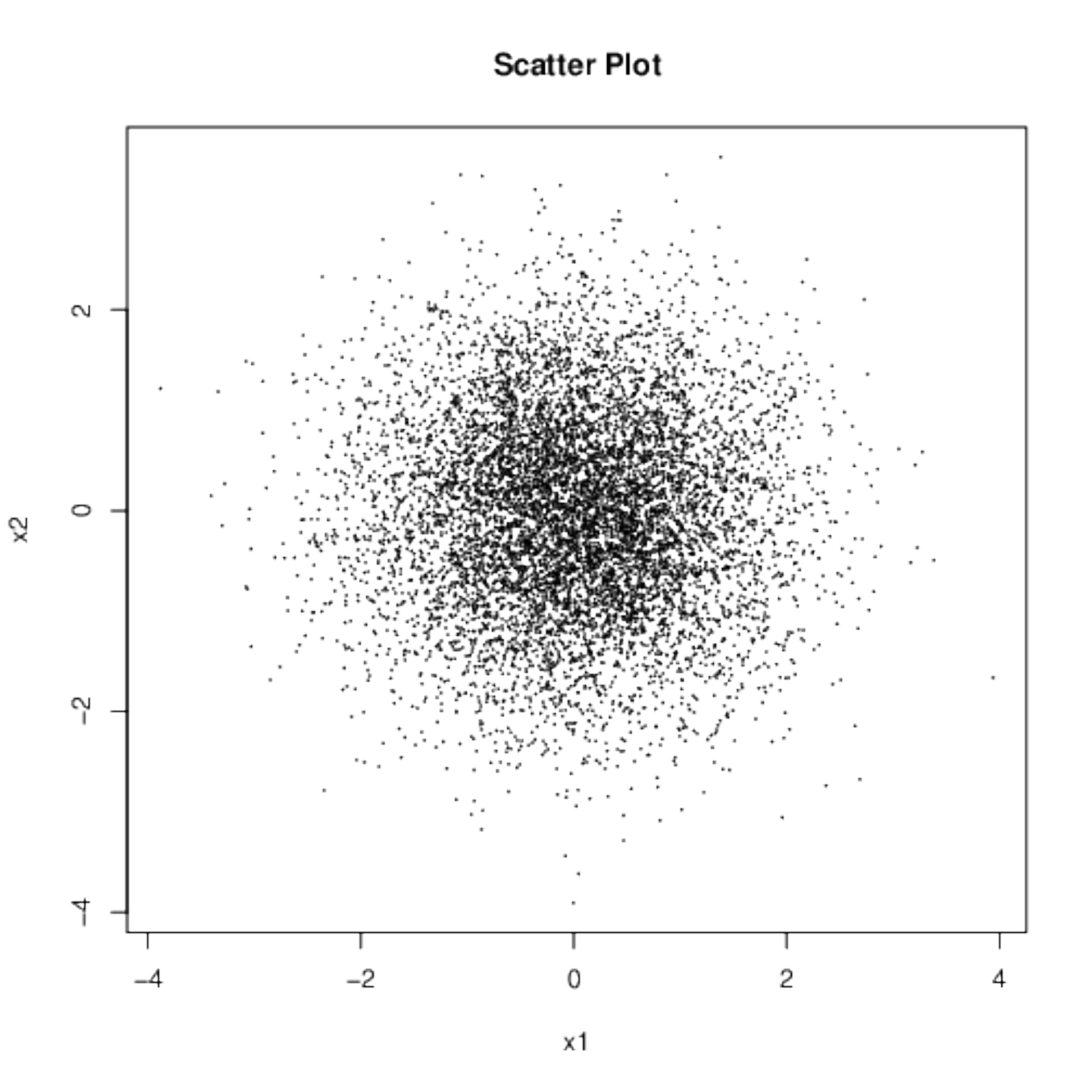} &
 \includegraphics[width=0.3\textwidth]{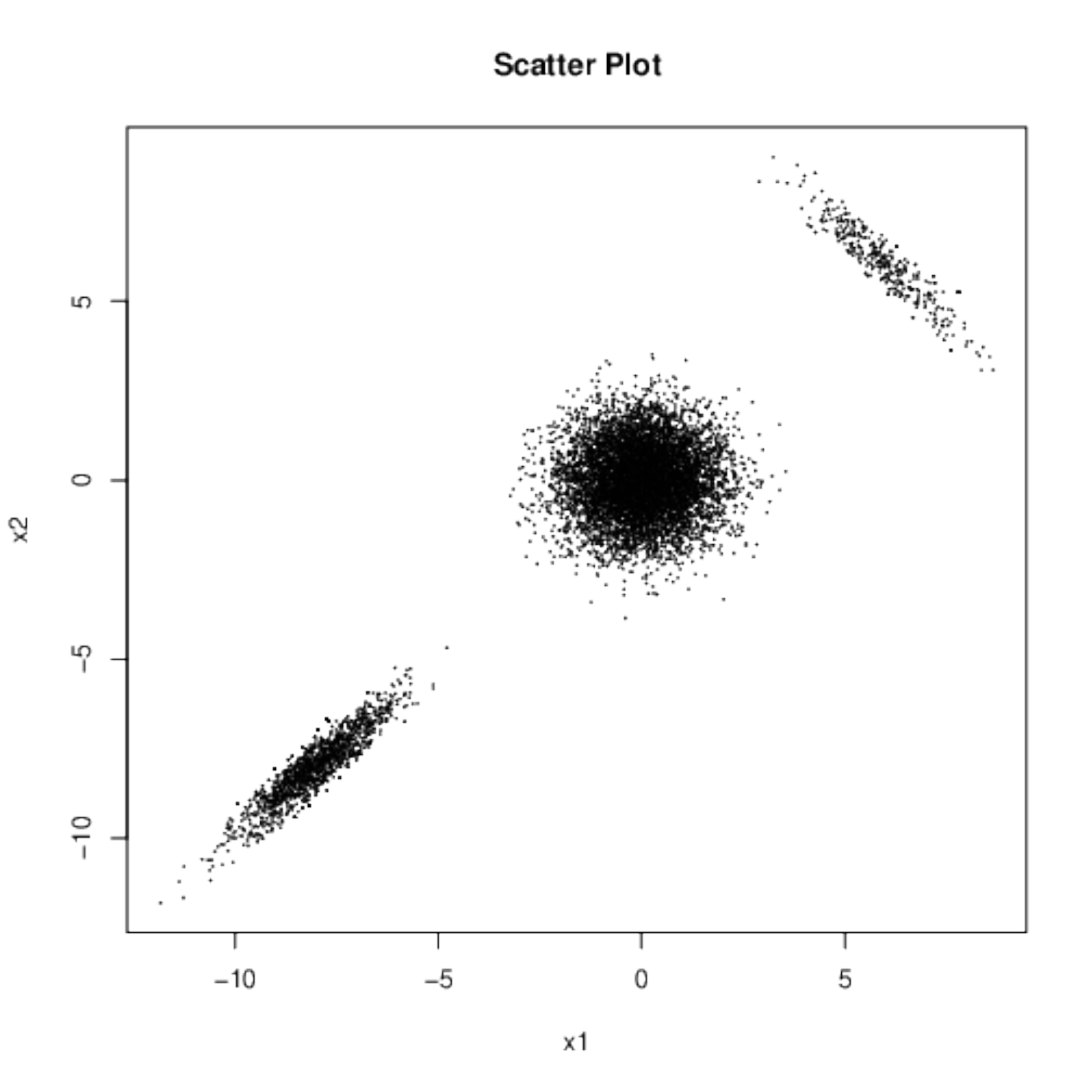} \\
\end{tabular}
\end{figure}

Figure \ref{fig:gmm}(a-c) show the position and momentum coordinates for the first 1000 iterations 
and the scatter plots of MCMC samples drawn by HMC and
Figure \ref{fig:gmm}(d-f) exhibits those of MCMC samples generated by SAHMC.
Scatter plots show that HMC cannot reach one of its three modes, 
whereas our SAHMC explores all parameter spaces.
Figure \ref{fig:gmm}(g-h) shows scatter plots of samples generated by 1st, 2nd
and 3rd run of NUTS. As we can see, NUTS fails to visit all three modes in a few
runs. In our simulation for Set 2, NUTS only visits all modes for five
independent runs out of total 10 runs.

The weight terms, $\theta^{(t)}$, which are updated adaptively based on the frequencies of chain visits,
help the SAHMC chains move more widely for both position and momentum coordinates
so that it can explore more sample spaces while maintaining the efficiencies of HMC. 
Therefore, our SAHMC method performs much better than HMC and NUTS when our target density has multi-mode. 

\subsection{High-dimensional Multimodal Distributions}

To investigate the performance of SAHMC in high-dimensional multimodal distributions, we consider an equal mixture of
eight $d$-dimensional Gaussians previously discussed by \cite{tak_repelling-attracting_2018}.
The target distribution is 
\begin{equation}
\label{eq:1}
\pi(x) \propto \sum_{j=1} ^{8} \exp \left( -0.5 (x-\mu_{j})^T(x-\mu_j) \right),
\end{equation}
where $x = (x_{1},x_{2},\ldots,x_d)^T$ and $\mu_j$ are determined by setting
their first three coordinates to the eight vertices of a cube with edge length
10. The remaining coordinates are filled by alternating $0$ and $10$:

\begin{align*}
  \mu_1 &= (10,10,10,0,10,0,10,\ldots,0,10),\\
  \mu_2 &= (0,0,0,10,0,10,0,\ldots,10,0),\\
  \mu_3 &= (10,0,10,0,10,0,10,\ldots,0,10), \\
  \mu_4 &= (0,10,10,0,10,0,10,\ldots,0,10),\\
  \mu_5 &= (0,0,10,0,10,0,10,\ldots,0,10),\\
  \mu_6 &= (0,10,0,10,0,10,0,\ldots,10,0),\\
  \mu_7 &= (10,0,0,10,0,10,0,\ldots,10,0),\\
  \mu_{8} &= (10,10,0,10,0,10,0,\ldots,10,0).\\
\end{align*}

We will demonstrate that SAHMC efficiently explore a multimodal
high-dimensional distribution in equation \ref{eq:1} with the five values of $d \in
\{3, 5, 7, 9, 11\}$. We compare SAHMC to vanilla HMC and NUTS.

For each $d$, HMC and SAHMC share the same the leapfrog step-size, $\epsilon$ and
leapfrog steps, $L$, which are listed in \ref{tab:mix8setup}. These values are chosen so that
the acceptance rate is between $0.4$ and $0.7$. 
We set the sample space was partitioned with equal energy bandwidth $\Delta u = 2$ into the following subregions: 
$E_1 = \{{\bf x}: -\log p({\bf x}) < u_{1} \}$, $E_2 = \{{\bf x}: 0 \leq -\log p({\bf x}) < u_{1}+2\}$, 
$\cdots$, and $E_{m} = \{{\bf x}: -\log p({\bf x}) > u_{1} + 2(m-1)\}$. The
values of $u_{1}$ and $m$ for each $d$ are listed in \ref{tab:mix8setup}.
The number of iterations are chosen so that each sampler produce approximately
equal computation time. Other configurations of samplers follow those in Section~\ref{sec:gauss-mixt-distr}.

\begin{table}[htbp]
\caption{\label{tab:mix8setup}
  High-dimensional multimodal distributions: Configurations of SAHMC and HMC 
  include the leapfrog step-size ($\epsilon$), leapfrog steps ($L$), the minimum
  energy level ($u_{1}$), and the number of energy partitions ($m$).}
\centering
\begin{tabular}{ccccc} \hline\hline
  d &
  $\epsilon$ & $L$ & $u_{1}$ & $m$ \\
  \hline

  3 & 0.9 & 1 & 8.0 & 6 \\
  5 & 0.25 & 3 & 8.0 & 10 \\
  7 & 0.25 & 3 & 8.0 & 14 \\
  9 & 0.25 & 3 & 8.0 & 18 \\
  11 & 0.25 & 3 & 8.0 & 22 \\
  % 13 & 0.25 & 3 & 8.0 & 26 \\
\end{tabular}
\end{table}

For each $d$, we run SAHMC ten times to obtain ten chains each
of length 1,000,000, discarding the first 200,000 iterations of each
chain as burn-in. As $d$ increases, SAHMC requires more evaluations because it is more difficult to find a proposal that increases the density in the forced uphill transition.

We use two measures to evaluate each method.
The first is $N_{dis}$, the average number of the unknown modes that are discovered by each
chain. A mode $\mu_{j}$ is tagged as uudiscovered'' when at least one sample has 
$\mu_j$ as the closest mode measured by the
Euclidean distance. 
The second is $F_{err} = \sum\nolimits_{i=1}^{10} \sum\nolimits_{j=1}^{8} |F_{i,j} -
1/8| / 80$, the average frequency error rate
\cite{tak_repelling-attracting_2018}, where $F_{i,j}$ is the proportion of
iterations in chain $i$ whose closest mode is $\mu_j$. 

Table~\ref{tab:mix8} summarizes the results, and shows that SAHMC is never worse
than NUTS in terms of $N_{dis}$ and $F_{err}$ regardless of dimension. HMC's
$F_{err}$ is only a bit smaller than that of SAHMC at $d=3$ but deteriorates much
faster than SAHMC's and NUTS's. Eight modes get more
isolated in terms of Euclidean distance as the dimension
Hence, there are more obstacles for samplers to travel across modes
for high-dimension. In high-dimensions, SAHMC performs
much efficiently by sampling across high energy barriers between multiple modes.

\begin{table}[htbp]
\caption{\label{tab:mix8}
  High-dimensional multimodal distributions: Comparison of SAHMC, HMC, and
  NUTS. The simulation results include the number of iterations; the burn-in size; $N_{dis}$ = the average number of modes
  discovered by each chain; and $F_{err} = \sum\nolimits_{i=1}^{10} \sum\nolimits_{j=1}^{8} |F_{i,j} -
  1/8| / 80$, where $F_{i,j}$ is the proportion of iterations in chain $i$ whose
  closest mode is $\mu_j$.}
\centering
\begin{tabular}{ccrclc} \hline\hline
  d &
  Kernel &
           \begin{tabular}{@{}c@{}}Length of a chain \\ (burn-in size)
            \end{tabular} & CPU time (s) & $N_{dis}$ & $F_{err}$ \\
  \hline
  \multirow{3}{*}{3}
 & SAHMC & 1,000,000 (200,000) & 2.98 & 8 & 0.0030 \\
 & HMC   & 1,000,000 (200,000) & 3.01 & 8 & 0.0024 \\
 & NUTS  & 555,000 (110,000)   & 3.72 & 8 & 0.0624 \\

  \hline
  \multirow{3}{*}{5}
 & SAHMC & 1,000,000 (200,000) & 5.24 & 8 & 0.0050 \\
 & HMC   & 1,000,000 (200,000) & 5.22 & 8 & 0.0246 \\
 & NUTS  & 687,500 (137,500)   & 5.83 & 8 & 0.0655 \\

  \hline
  \multirow{3}{*}{7}
 & SAHMC & 1,000,000 (200,000) & 5.59 & 8   & 0.0081 \\
 & HMC   & 1,000,000 (200,000) & 5.63 & 4.4 & 0.1248 \\
 & NUTS  & 500,000 (100,000)   & 5.90 & 8   & 0.0955 \\

  \hline
  \multirow{3}{*}{9}
    & SAHMC & 1,000,000 (200,000) & 5.87 & 8 & 0.0265 \\
 & HMC   & 1,000,000 (200,000) & 5.83 & 4   & 0.1250 \\
    & NUTS  & 375,000 (\hphantom{0}75,000)    & 6.10 & 8   & 0.1036 \\

  \hline
  \multirow{3}{*}{11}
 & SAHMC & 1,000,000 (200,000) & 6.11 & 8 & 0.0431 \\
 & HMC   & 1,000,000 (200,000) & 6.13 & 4 & 0.1250 \\
    & NUTS  & 350,000 (\hphantom{0}70,000)    & 7.26 & 8 & 0.1082 \\     

\end{tabular}
\end{table}

\subsection{Sensor Network Localization}

We illustrate the advantage of SAHMC using a sensor network localization problem previously discussed by
\cite{tak_repelling-attracting_2018,ahn_distributed_2013,ihler_nonparametric_2005}.
This problem is known to produce a high-dimensional and multimodal joint posterior distribution.
Following the experiment
setting in \cite{tak_repelling-attracting_2018}, we assume N sensors are scattered in a planar
region with 2d locations denoted by $\{x_t\}_{t=1}^N$. The distance
between a pair of sensors $(x_{t},x_u)$ is observed with a probability
$\exp(- 0.5 ||x_t - x_u||^2 / R^2 )$, and the observed distance between $x_{t}$
and $x_{u}$, denoted by $Y_{t,u}$, 
follows a Gaussian distribution with mean $||x_t - x_u||$ and standard deviation
$\sigma$. Independent bivariate Gaussian prior
distributions with mean $(0, 0)$ and covariance matrix $10^2 \times I_2$ are assumed
for $x_{t}$'s.  Given a set of observations
$\{y_{t,u}\}$,  a typical task is to infer the posterior distribution of all the
sensor locations. Following \cite{tak_repelling-attracting_2018}, we choose $N =
4$, $R = 0.3$, $\sigma = 0.02$ and add two additional sensors with known
locations. The locations of the 4 sensors form a
multimodal distribution of 8 dimensions.

For both HMC and SAHMC, we set the leapfrog step-size, $\epsilon = 0.02$, and
leapfrog steps, $L = 3$. For SAHMC, the sample space was partitioned with equal
energy bandwidth $\Delta u = 2$, the minimum energy level $u_{1} = -4$ and the
number of partitions $m = 19$. We implemented SAHMC and HMC for 2,000,000 iterations with the first 400,000
as burn-in, resulting in 106 seconds computation time. For a fair comparison in
terms of computation time, we
implemented NUTS for 400,000 iterations with the first 80,000 as burn-in,
resulting in 102 seconds computation time. All other configurations of samplers
follow those described in Section~\ref{sec:gauss-mixt-distr}.

\begin{figure}[htbp]
  \caption{\label{fig:sensor}
    Scatterplots of the posterior sample of each sensor location obtained by
    SAHMC, HMC and NUTS. The coordinates of the true location are denoted by dashed lines.}
  \centering
  \begin{tabular}{ccc}
    \includegraphics[width=0.3\textwidth]{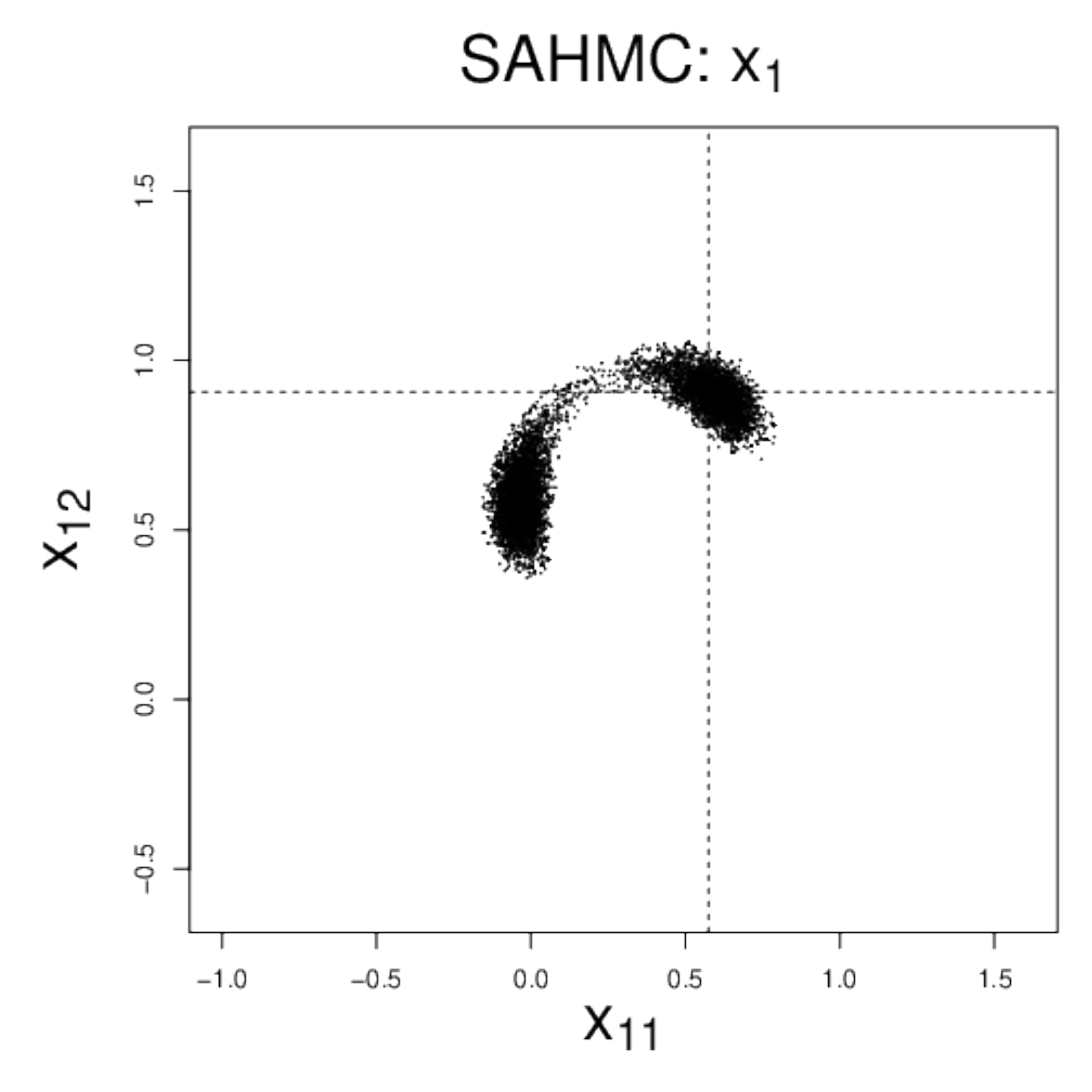} &
                                                                    \includegraphics[width=0.3\textwidth]{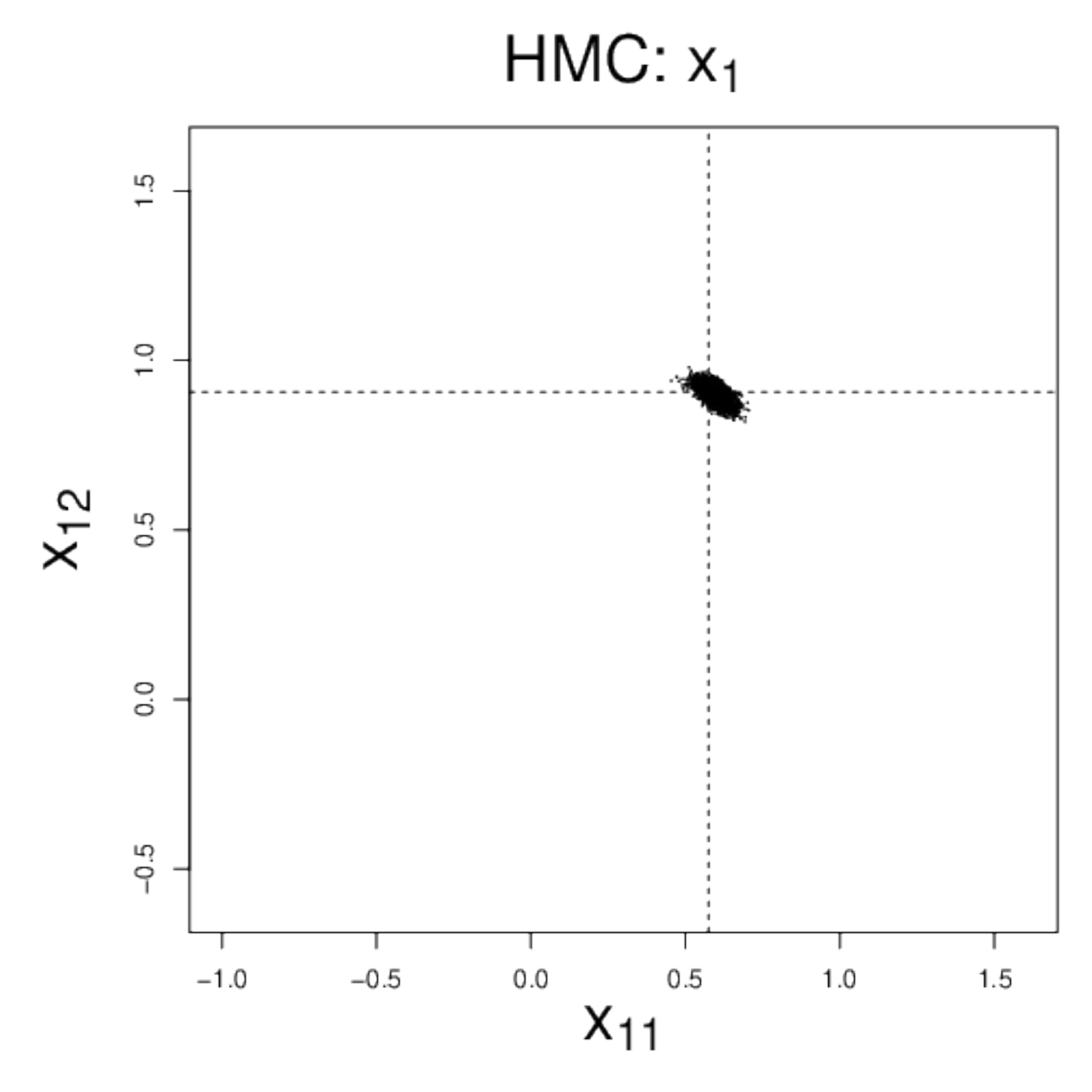} &
                                                                                                                                   \includegraphics[width=0.3\textwidth]{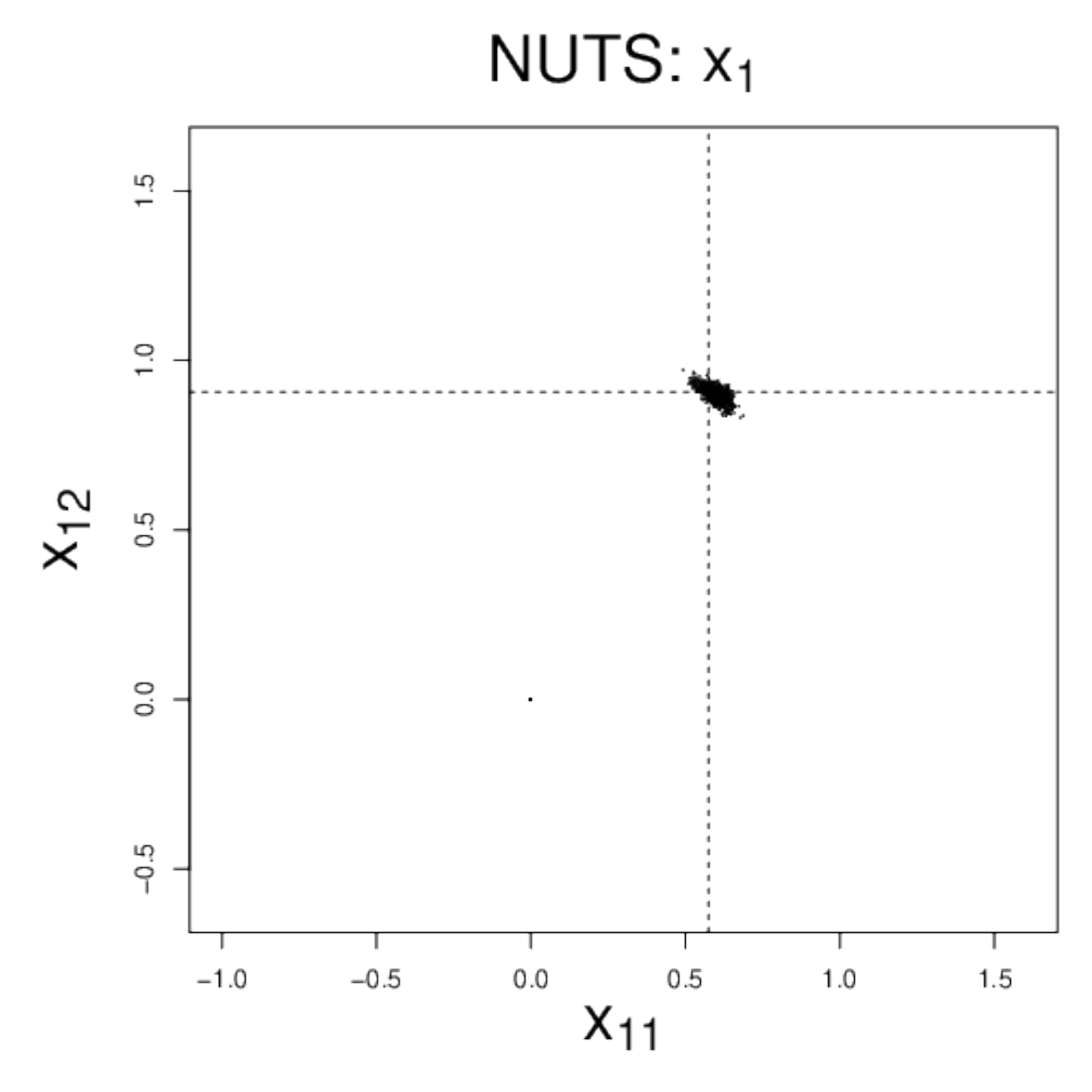} \\
    \includegraphics[width=0.3\textwidth]{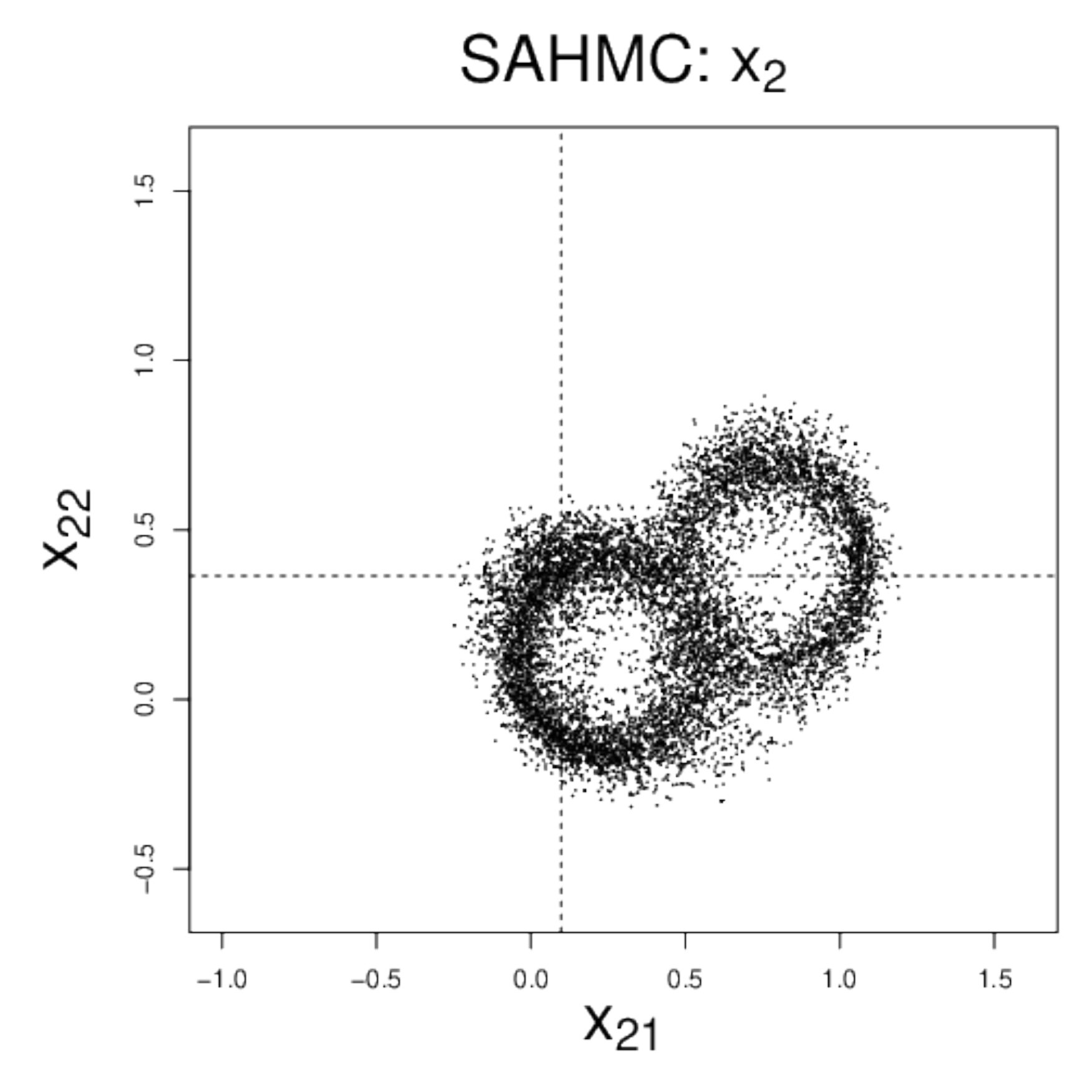} &
                                                             \includegraphics[width=0.3\textwidth]{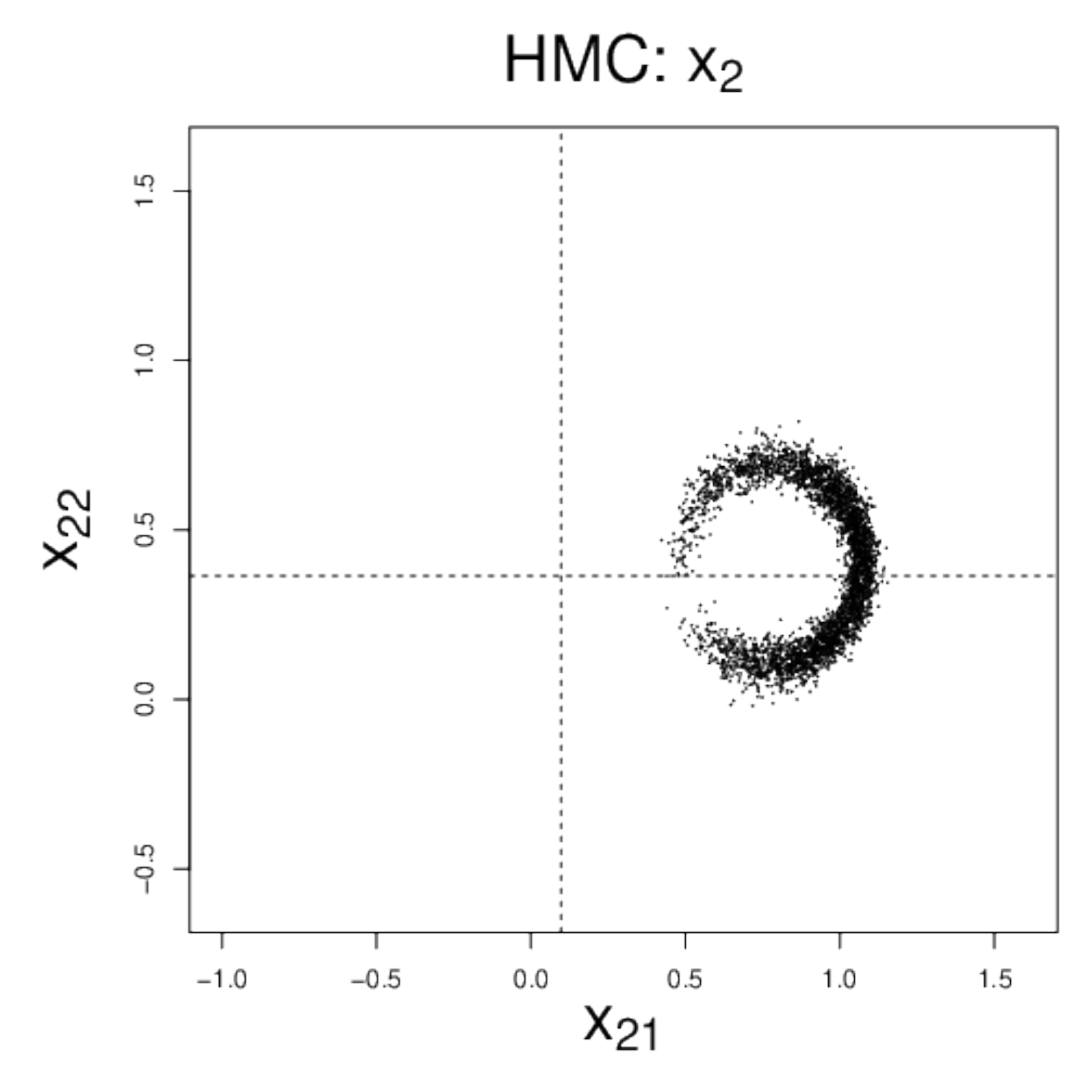} &
                                                                                                                            \includegraphics[width=0.3\textwidth]{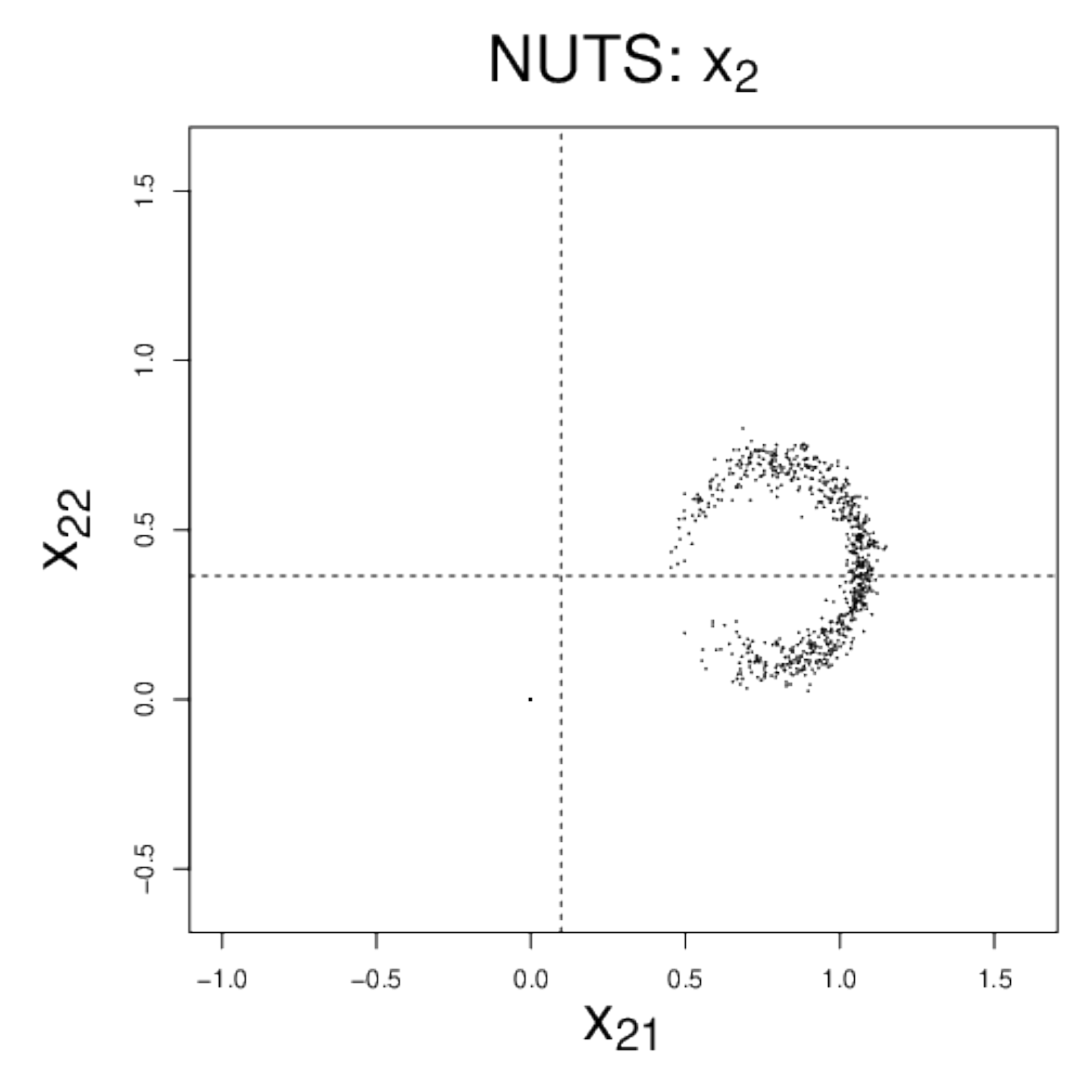} \\
    \includegraphics[width=0.3\textwidth]{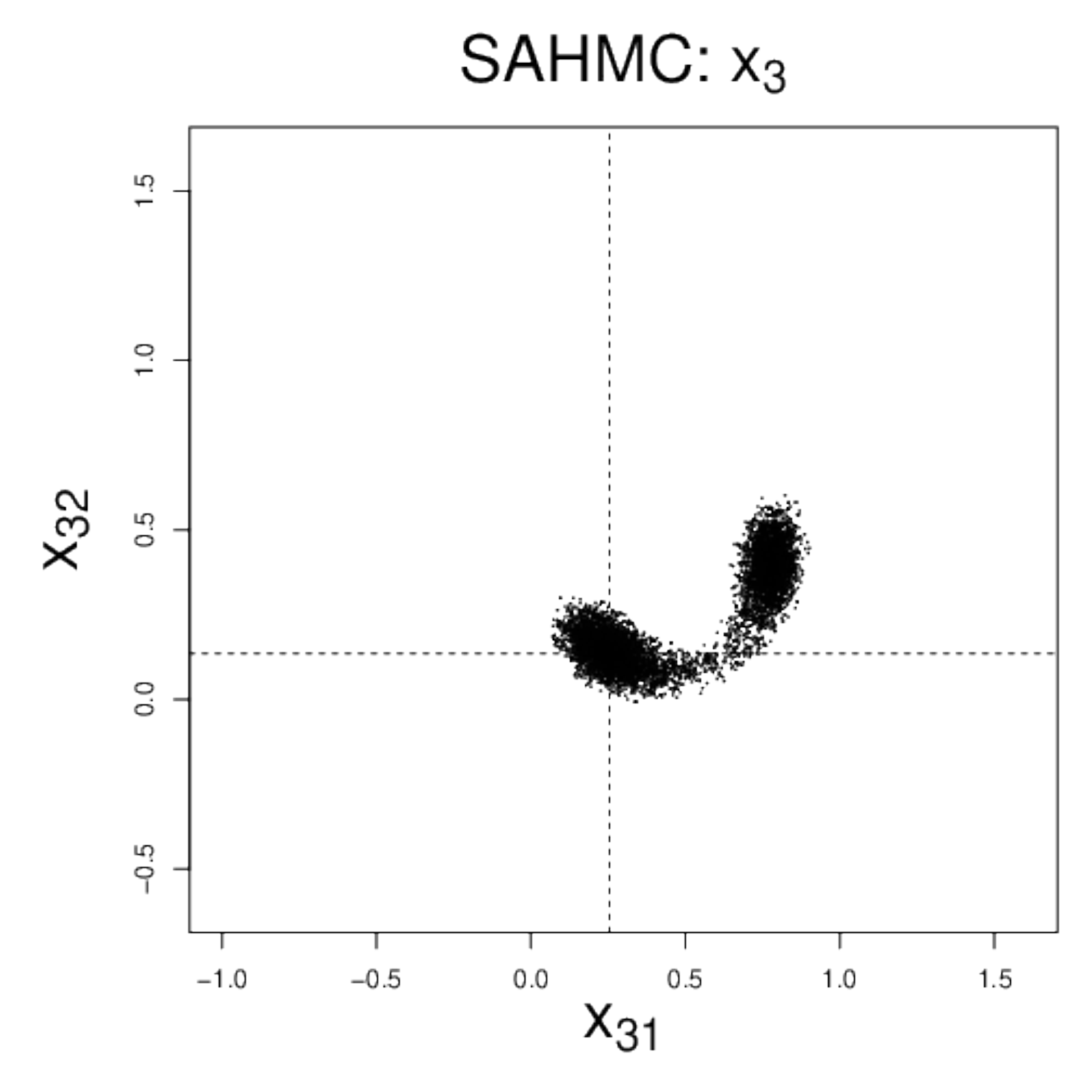} &
                                                             \includegraphics[width=0.3\textwidth]{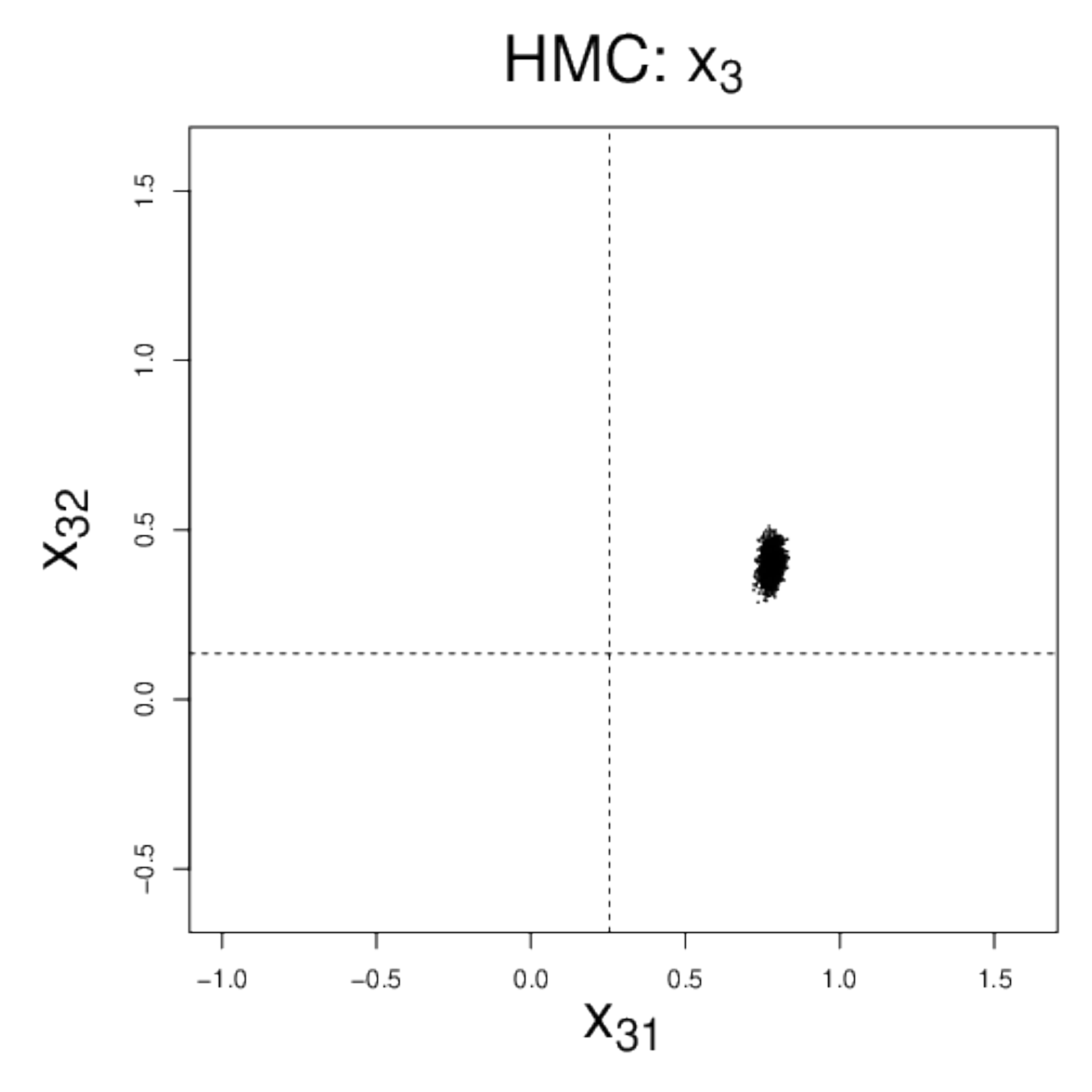} &
                                                                                                                            \includegraphics[width=0.3\textwidth]{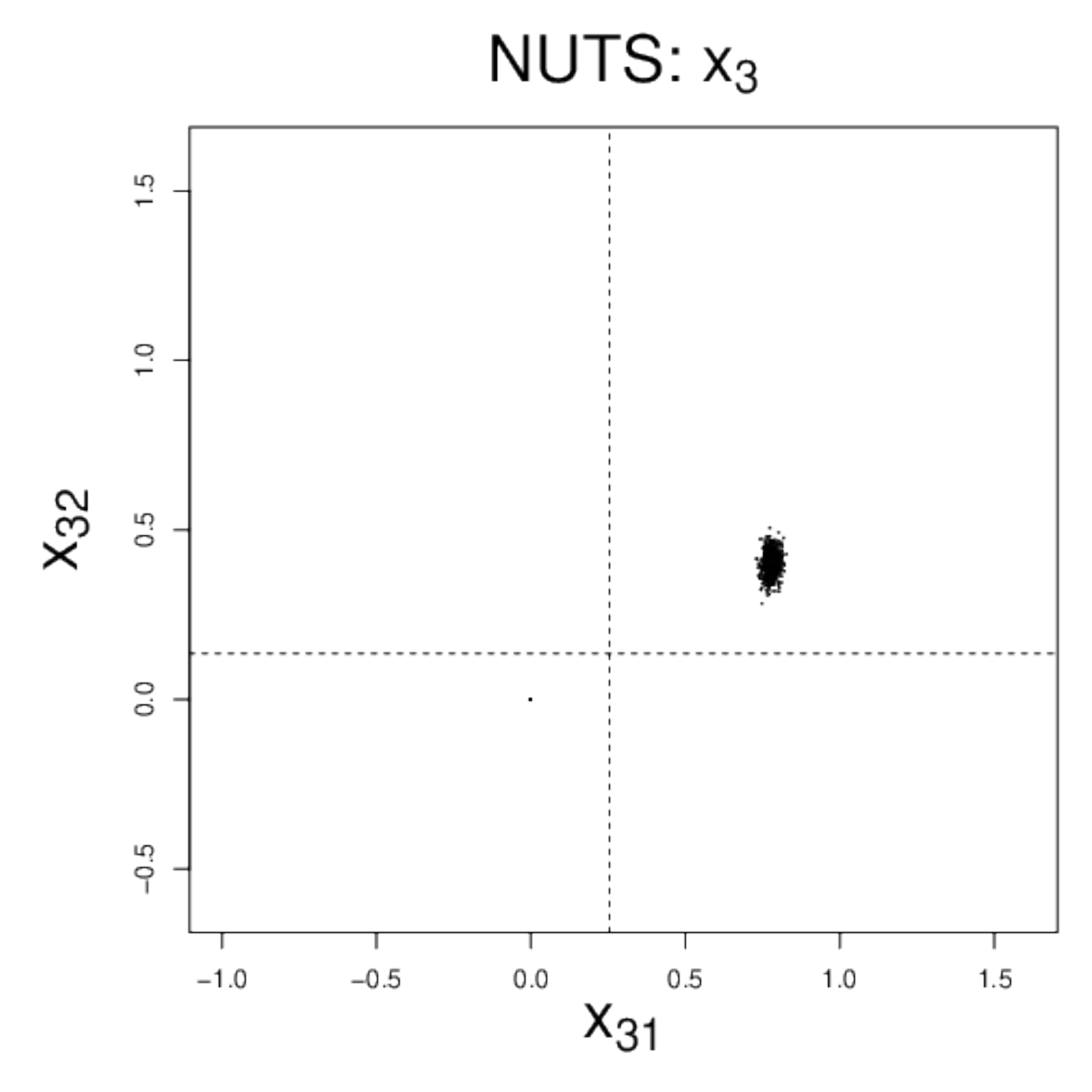} \\
    \includegraphics[width=0.3\textwidth]{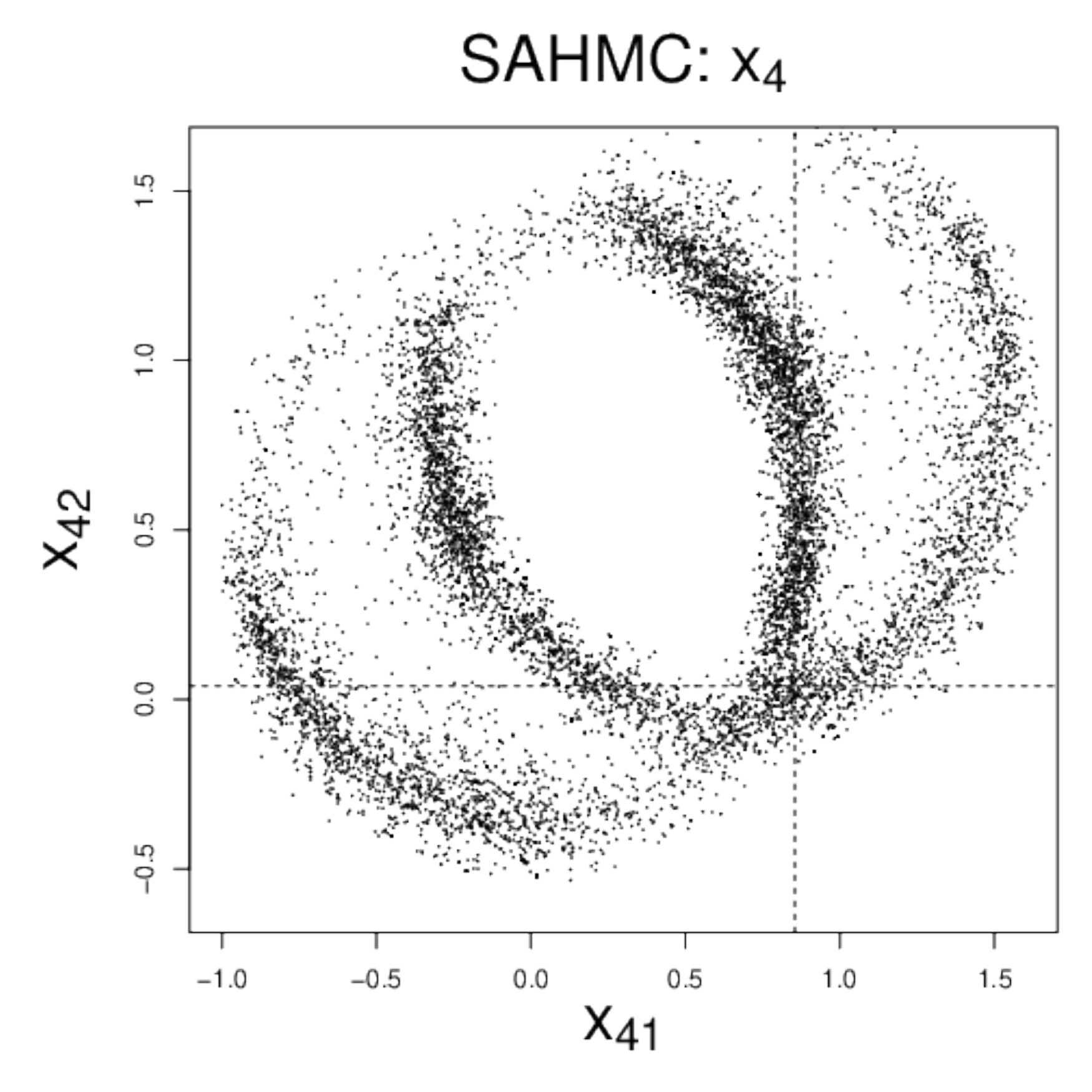} &
                                                             \includegraphics[width=0.3\textwidth]{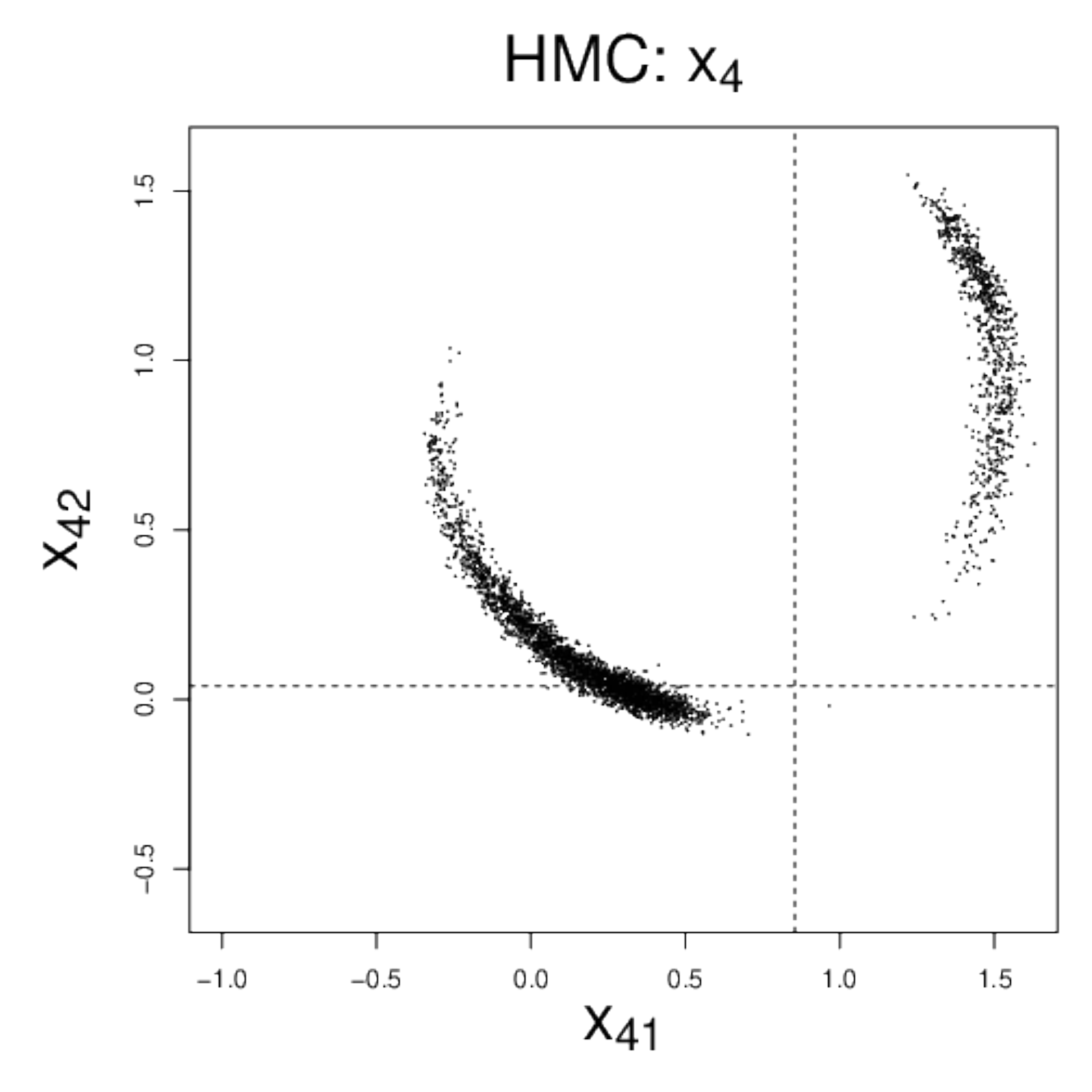} &
                                                                                                                            \includegraphics[width=0.3\textwidth]{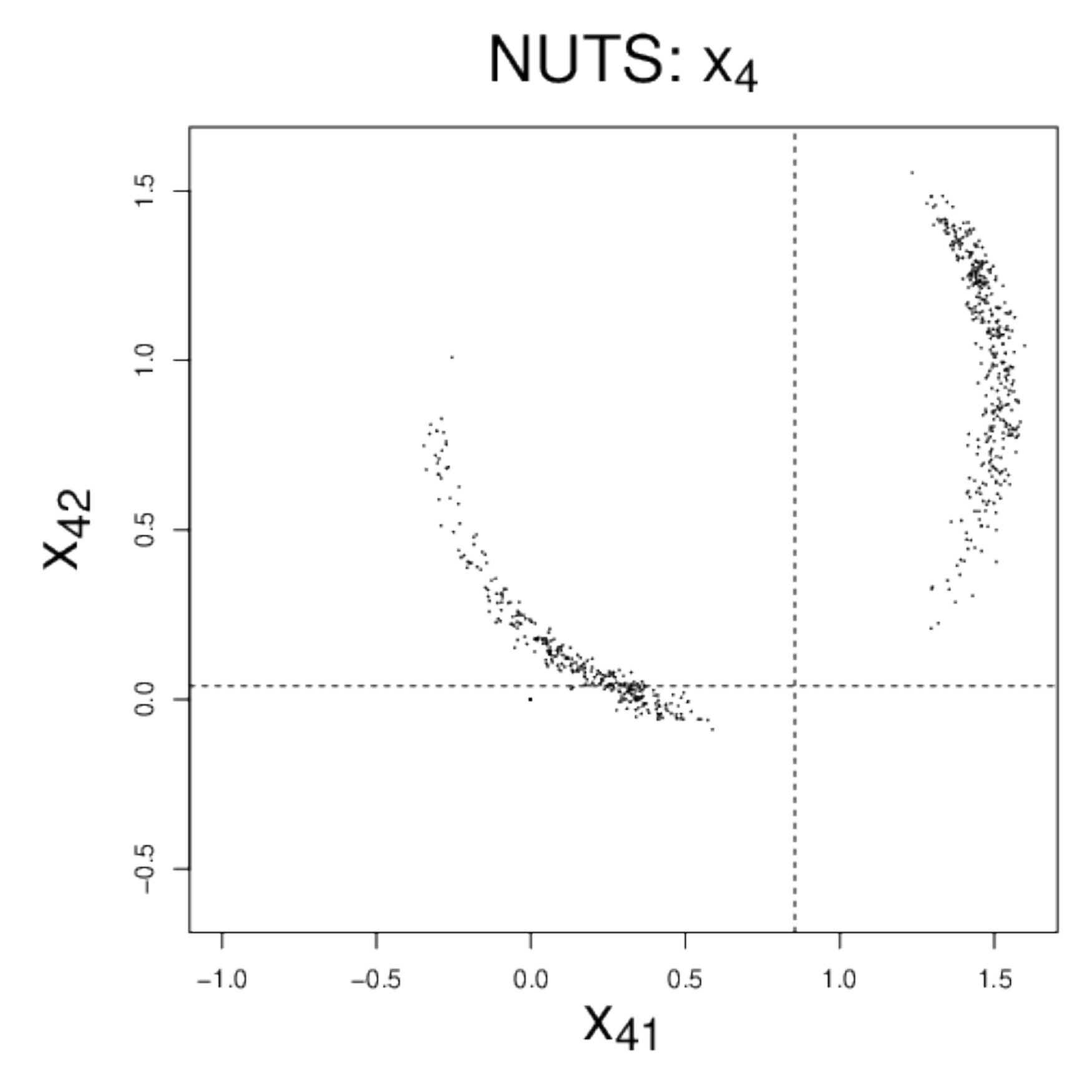} \\

  \end{tabular}
\end{figure}

Figure~\ref{fig:sensor} shows scatter plots of the posterior samples of each sensor
location obtained by the three samplers. The dashed lines
indicate the coordinates of the true location. We can see that HMC and NUTS fail
to visit one of the modes; whereas, SAHMC frequently visits this mode and
generate enough samples from it.

\section{An Application to Neural Networks}

%Let $L$ denote the number of hidden layers, $N_l$ denote the number of nodes 
%in layer $l\in\{1,\dots,L\}$. 
%Let $N = \sum_{l=1}^L N_l$ denote the total number of nodes, 
%and ${\bf x}=\{x^{(1)}, \dots, x^{(p)} \}$ denote the input of neural network 
%for $x^{(j)}\in\mathbb{R}^n$ and sample size $n$. 
%We denote the output of the $j$-th node of layer $l$ by $z_j^{l}$, 
%and the weight of the edge connecting node $i$ in layer $l$ 
%and node $j$ in layer $l+1$ by $w_{i,j}^l$ for $i\in\{1,\dots,N_l\}$ and $j\in\{1,\dots,N_{l+1}\}$. 
%Let $b_j^l$ denote the bias of the node $j$ in layer $l$. 
%Then, an feed-forwarding neural network is defined by a feedforward network 
%\citep{Schmidhuber:12} that is expressible as 
%\begin{eqnarray*}
%z_j^{l+1} = \sigma\left(  \sum^{N_l}_{i=1} w_{i,j}^lz_i^l + b_j^{l+1} \right),
%\end{eqnarray*}
%for $l\in\{1,\dots,L-1\}$ and $j\in\{1,\dots,N_{l+1}\}$, with
%\begin{eqnarray*}
%\mbox{input layer}&:& z_j^1 = \sigma \left(  \sum^{p}_{i=1} w_{i,j}^0x^{(i)} + b_j^{1} \right), \:\:\:j\in\{1,\dots,N_1\}, \\
%\mbox{output layer}&:& f({\bf x}) = \sigma\left(  \sum^{N_L}_{i=1} w_{i,j}^Lz_i^L + b_j^{L+1} \right),
%\end{eqnarray*}
%where $\sigma(\cdot)$ denotes the activation function, and we model the function that involves 
%in prediction or classification of data by the function of the feedforward neural network $f$. 

Feed-forward neural networks, which are also known as multiple layer perceptrons (MLP), 
are one of well-known models in machine learning community \citep{Schmidhuber:12}. 
Given a group of connection weights ${\bf z} = (\alpha, \beta)$, the MLP can be written as
\begin{equation}\label{eq:MLP}
 f\Big({\bf x}_i \mid {\bf z}\Big) = \varphi\bigg(\alpha_0 %+ \sum_{j=1}^p \gamma_j x_{ij} 
 + \sum_{k = 1}^N \alpha_k\varphi\Big(\beta_{k0} + \sum_{j=1}^p \beta_{kj} x_{ij}\Big)\bigg),
\end{equation}
where $N$ is the number of hidden units, $p$ is the number of input units, 
${\bf x}_i = \big(x_{i1}, \cdots, x_{ip}\big)$ is the $i$-th input patterns, and 
$\alpha_k$, and $\beta_{kj}$ are the weights on the connections 
from the $k$-th hidden unit to the output unit,  
from the $j$-th input unit to the $k$-th hidden unit, respectively. 
The function $\varphi$ is the activation function of the hidden and output units. 
Popular choices of $\varphi(\cdot)$ include the sigmoid function, the hyperbolic tangent function, 
and the rectified linear unit (ReLU).
%$\alpha_k$, $\beta_{kj}$, and $\gamma_j$ are the weights on the connections from the $k$-th hidden unit 
%to the output unit,  
%from the $j$-th input unit to the $k$-th hidden unit, 
%and from the $j$-th input unit to the output unit, respectively.

\begin{figure}[htbp]
\centering
\begin{tabular}{cc}
(a) SAHMC & (b) HMC\\
\includegraphics[width=0.4\textwidth]{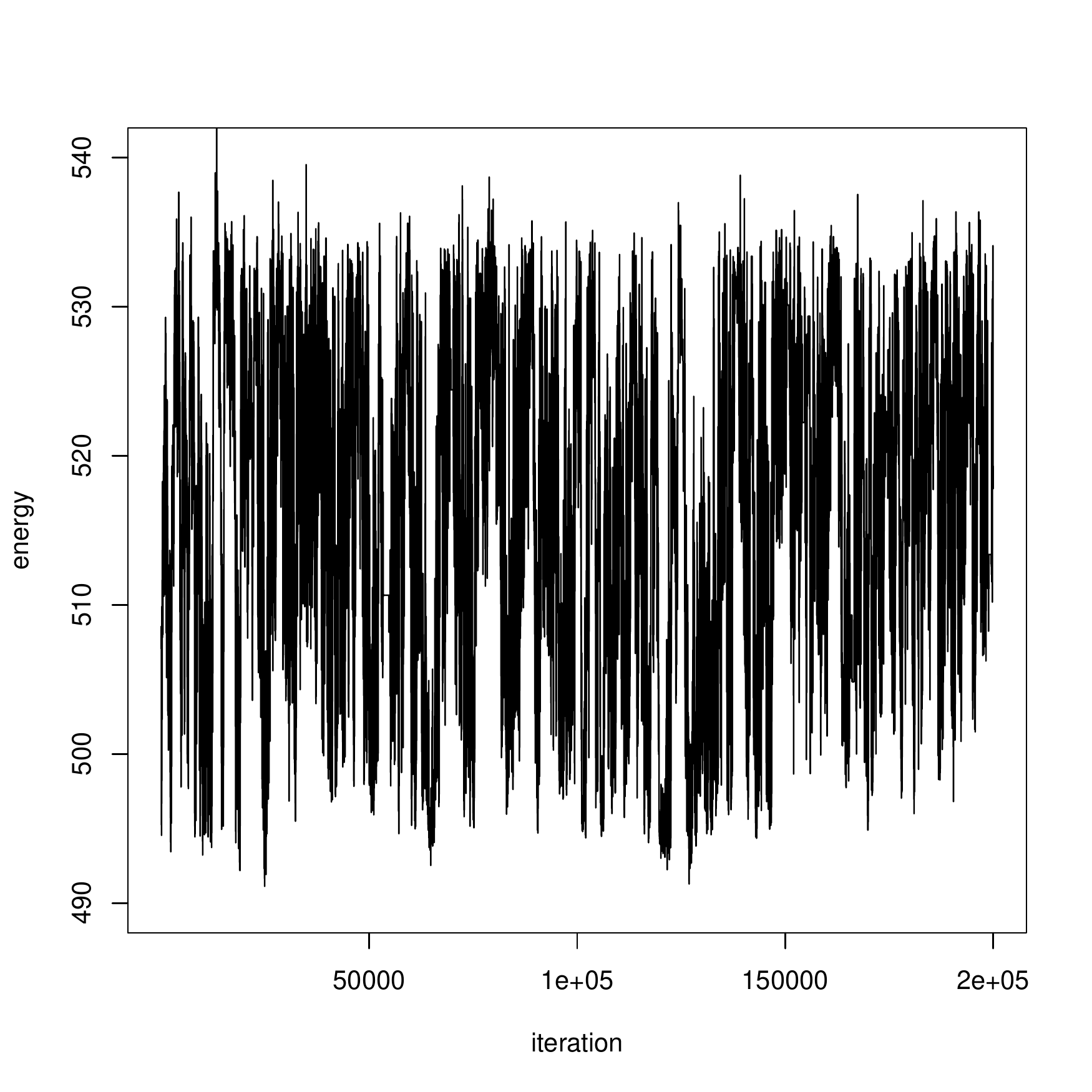} &
\includegraphics[width=0.4\textwidth]{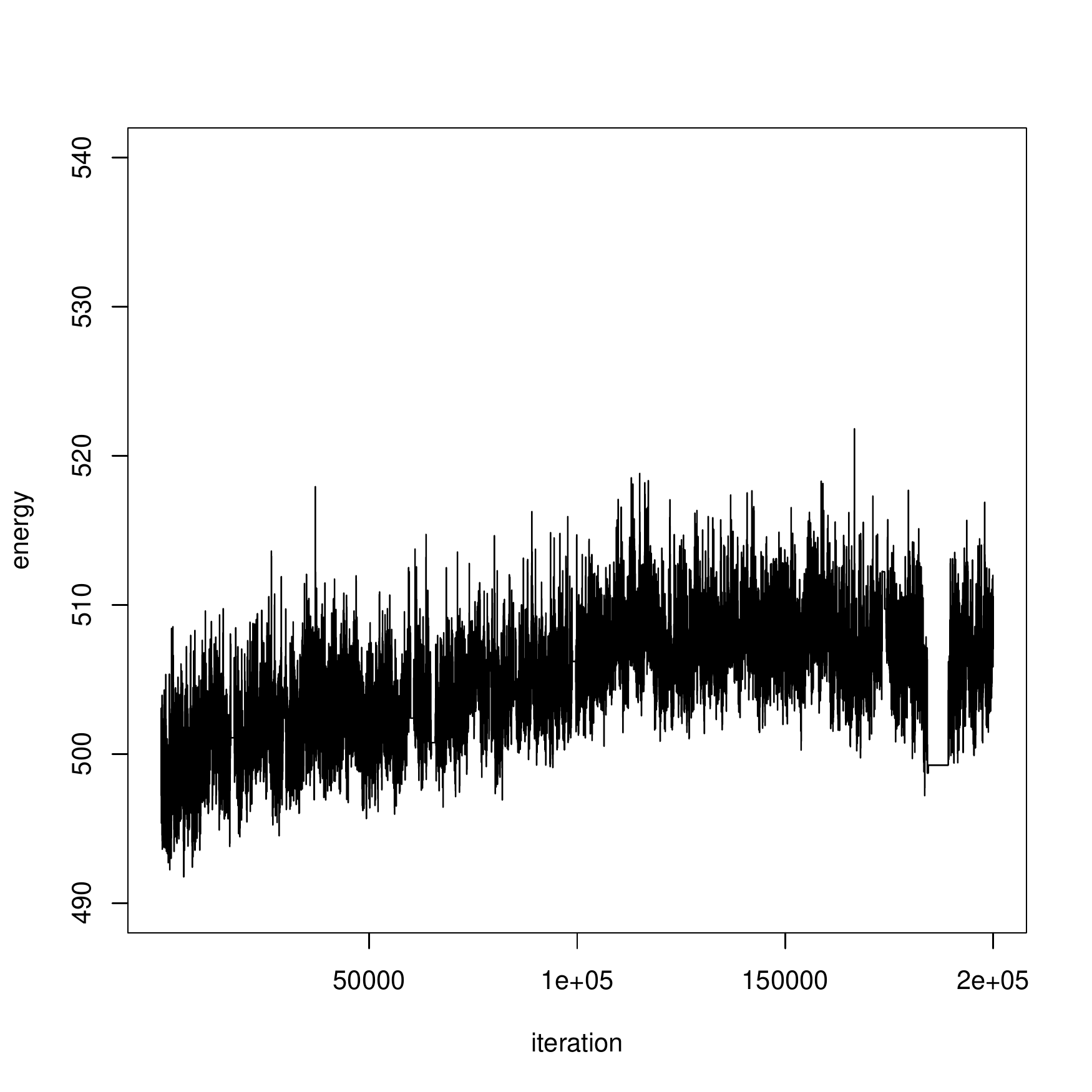}\\
\end{tabular}
\caption{\label{fig:trace} An example of trace plots.}
\end{figure}

%By using the neural network on $f$, we consider the following nonparametric regression models:
%\begin{eqnarray*}
% y_i = f(x_i) + \epsilon_i	
%\end{eqnarray*}
%where $\epsilon_i\sim N(0,\eta)$ with the error variance $\eta$ for $i=1,\dots,p$.

There have been multiple studies regarding computational aspects of  Bayesian neural network models via MCMC algorithms 
\citep{andrieu2000reversible, lampinen2001bayesian}%, muller1998issues}, 
but their practical performance were questioned 
due to the highly correlated parameters on posterior space. 
Alternatively, in \citet{neal2012bayesian} the HMC was used to sample the weight parameters, 
 improving the convergence of the MCMC chain. 
However, the highly multimodal nature of the posterior distribution of the MLP  
still hinders the practical implementation of neural network models. 
To solve this issue, we apply the SAHMC to neural network models, 
and consider simulated datasets to examine the capacity of the SAHMC to efficiently 
explore the multimodal posterior space. 

\paragraph*{A Simulation Study}
We consider a simple regression settings with one predictor for one-layered feedforward neural network, 
and we generate the data from $y_i = f_0(x_i) + \epsilon_i$, 
where the true function $f_0(x) = 3\varphi(x - 1.5) - \varphi(x + 1) - 3\varphi(x-1) + 2\varphi(x)$ 
with the ReLU function $\varphi(x) = \max\{0,x\}$, and $\epsilon_i\sim N(0,1)$ for $i=1, \dots, n$. 
We independently replicates 200 simulated data sets 
and use SAHMC and HMC to sample from the posterior distribution of the connection weights.

For both HMC and SAHMC, we set the leapfrog step-size, $\epsilon = 0.005$, and leapfrog steps, $L = 25$.
To run SAHMC, we set the sample space ${\cal X} = [-10^{100}, 10^{100}]^2$ to be compact and
it was partitioned with equal energy bandwidth $\Delta u = 2$ into the following subregions: 
$E_1 = \{{\bf x}: -\log p({\bf x}) < 460\}$, $E_2 = \{{\bf x}: 460 \leq -\log p({\bf x}) < 462\}$, 
$\cdots$, and $E_{51} = \{{\bf x}: -\log p({\bf x}) > 560\}$.
Additionally, we set $t_0 = 1000$ and the desired sampling distribution, $\boldsymbol{\pi}$ to be uniform for SAHMC. 
Both HMC and SAHMC were independently run and 
each run consists of 55,000 iterations, where the first 5,000 iterations were discarded as a burn-in process. 
All initial points are randomly selected for all simulations. 

To measure the performance of each procedure, 
we consider a {\it Posterior risk} of $f(x|{\bf z})$, that is $\sum_{i=1}^n \mathbb{E
}_{\bf z\mid y}\left[ \{f_0(x_i) - f(x_i\mid {\bf z})\}^2 \right]$, where $\mathbb{E
}_{\bf z\mid y}$ is the expectation operator with respect to the posterior distribution of ${\bf z}$, and we also consider the ESS. 
We also report the minimum energy found by the SAHMC and the HMC and the proportion of the cases 
where the SAHMC procedure finds the smaller energy region than the energy found by the other procedure. 
The results are averaged over 200 replicated data sets.  

\begin{table}[htbp]
\centering
\begin{tabular}{lcc} \hline
%\multicolumn{6}{l}{{\bf Set 1: a = -6, b = 4}}\\
Method & Posterior Loss & ESS (min, med, max)  \\ \hline
 HMC &  0.112 & (4.1, 40.3, 282.7)  \\ 
 SAHMC   & 0.077 & (8.3,  184.9, 335.8)   \\ 
 \hline 
\end{tabular}
\caption{\label{tab:ANN}
Results of the Simulation Study: Comparison of SAHMC and HMC}
\end{table}

Table \ref{tab:ANN} summarizes the performance of the SAHMC and the HMC 
shows that the posterior expectation of $L_2$ loss from the regression function evaluated 
by the SAHMC achieves a smaller than that from the HMC. 
The median ESS of the SAHMC is about 4.5 times larger than that of the HMC.

In Figure \ref{fig:trace}, we provide an example of the trace plot of a SAHMC chain 
and a HMC chain for a simulated data set. 
This example illustrates how different the Markov chains of SAHMC and HMC are. 
The SAHMC chain explores all energy level between 503 and 571, 
while the HMC chain searches only narrower energy level between 509 to 541. 
This shows the capacity of SAHMC in escaping local maxima of the posterior distribution 
and exploring wider region of the posterior space.     

\paragraph*{Pima Indians Diabetes Classification}
We consider a real data set that contains 768 records of female Pima Indians, 
characterized by eight physiological predictors and the presence of diabetes \citep{smith1988using}. 
We model the data by an artificial neural network (ANN) of a single layer equipped with 25 hidden nodes. 
The ANN is trained using SAHMC and HMC on a randomly selected $90\%$ of samples, 
and the out-sample prediction error was evaluated over $10\%$ of the other samples. 
We replicates this procedure 100 times and report the test error, ESS, and the average of minimum energy values found by each procedure. 
All other algorithm settings are same except the energy partitions: 
$E_1 = \{{\bf x}: -\log p({\bf x}) < 290\}$, $E_2 = \{{\bf x}: 290 \leq -\log p({\bf x}) < 292\}$, 
$\cdots$, and $E_{36} = \{{\bf x}: -\log p({\bf x}) > 360\}$.

\begin{table}[htbp]
\centering
\begin{tabular}{lccc} \hline
%\multicolumn{6}{l}{{\bf Set 1: a = -6, b = 4}}\\
Method & Test Error & ESS (min, med, max) & min.Energy  \\ \hline
 HMC & 0.383  & (1.7,4.0,24.9)  &  294.90\\ 
SAHMC   & 0.265  &  (4.4,15.0,34.5)  &  289.13 \\ 
 \hline 
\end{tabular}
\caption{\label{tab:real}Pima Indians Diabetes Data Set}
\end{table}
Table \ref{tab:real} shows that the SAHMC algorithm collects more effective samples and achieves a smaller test error  compared to the HMC. The average of the minimum energy searched by the SAHMC is also 5.77 lower  than that found by the HMC.

\section{Concluding Remarks}

In this paper, we propose a new algorithm which generates samples from multimodal density under the HMC framework. 
Because SAHMC can adaptively lower the energy barrier, 
our proposed algorithm, SAHMC can explore the rugged energy space efficiently. 
We compare the results of SAHMC with those of HMC and NUTS and show that SAHMC works more efficiently 
when there exists multiple modes in our target density, especially in high dimension.

RMHMC \citep{Girolami:11} can be easily combined with SAHMC by replacing the fixed mass matrix,
$\bf M$ with a position dependent expected Fisher information
matrix $\bf M(\bf x)$. We have implemented this algorithm and observed some
performance gain in terms of ESS per iteration.
However, $\bf M(\bf x)$ should be updated at each iteration using the fixed
point iteration, which poses a computational bottleneck. As a result, its
performance per CPU time is not as good as that of SAHMC.

One of the main issues that arise when HMC is used, is parameter tuning of the
algorithm (e.g. the mass matrix, $\bf{M}$, $\epsilon$ and $L$ in the leapfrog
integrator). A choice of these parameters is essential for good convergence of
the sampler. Combining HMC with SAMC, our proposed sampler shows its ability to
fully explore complex target distributions without much efforts to tuning these
parameters to make HMC work. It should be noted that HMC samplers using the same
parameters show poor convergence.

One difficulty in the application of SAHMC is how to set up the boundary of sample space partition. 
One approach we can take to overcome this difficulty is running our SAHMC with two stages. 
At the first stage, we run HMC with a few hundreds iterations, and then,
run SAHMC with the range of the sample space determined by the results of the first stages.

\section*{Acknowledgement} 

Ick Hoon Jin was partially supported by the Yonsei University Research Fund of 2019-22-0210 and by Basic Science Research Program through the National Research Foundation of Korea (NRF 2020R1A2C1A01009881). 

\bibliography{reference}

\end{document}